\documentclass[fleqn,usenatbib,a4paper]{mnras}

\usepackage{newtxtext,newtxmath}

\usepackage[T1]{fontenc}
\usepackage{ae,aecompl}

\usepackage{graphicx}	
\usepackage{amsmath}	
\usepackage{amssymb}	
\usepackage{mathtools}
\usepackage{hyperref}
\usepackage{xspace}	
\usepackage{euscript}
\usepackage{pdflscape}
\usepackage{afterpage}
\usepackage{placeins}
\usepackage{xcolor}
\usepackage{mleftright}
\usepackage[autostyle]{csquotes}
\usepackage{booktabs}

\newcommand{\edits}[1]{{#1}}
\newcommand{\gaia}{{\it Gaia}\xspace}
\newcommand{\hipparcos}{{Hipparcos} \xspace}
\newcommand\footnoteref[1]{\protected@xdef\@thefnmark{\ref{#1}}\@footnotemark}

\title[When and where were \textit{Gaia}'s eyes on the sky?]{Completeness of the \textit{Gaia}-verse I: when and where were \textit{Gaia}'s eyes on the sky during DR2?}

\author[D. Boubert, A. Everall, and B. Holl]{ 
	Douglas Boubert$^{1}$\thanks{E-mail: douglas.boubert@magd.ox.ac.uk},
	Andrew Everall$^{2}$
	 \edits{and Berry Holl$^{3,4}$}
	\\
	$^{1}$Magdalen College, University of Oxford, High Street, Oxford OX1 4AU, UK\\
	$^{2}$Institute of Astronomy, University of Cambridge, Madingley Road, Cambridge CB3 0HA, UK\\
	$^{3}$ Department of Astronomy, University of Geneva, Ch. des Maillettes 51, CH-1290 Versoix, Switzerland \\
    $^{4}$ Department of Astronomy, University of Geneva, Ch. d’Ecogia 16, CH-1290 Versoix, Switzerland \\
}

\date{Accepted XXX. Received YYY; in original form ZZZ}

\pubyear{}

\begin{document}
\label{firstpage}
\pagerange{\pageref{firstpage}--\pageref{lastpage}}
\maketitle

\begin{abstract}
	The \gaia space mission is crafting revolutionary astrometric, photometric and spectroscopic catalogues that will allow us to map our Galaxy, but only if we know the completeness of this \gaia-verse of catalogues: what stars does it contain and what stars is it missing? We argue that the completeness is driven by \gaia's spinning-and-precessing scanning law and will apply this principle to the \gaia-verse over this series. We take a first step by identifying the \edits{periods in time that did not contribute any measurements to \gaia DR2}; these gaps create ribbons of incompleteness across the sky that will bias any study that ignores them\edits{, although some of these gaps may be filled in future data releases}. \edits{Our first approach was to use the variable star photometry to identify the 94 gaps longer than 1\% of a day.} Our second approach was to predict the number of observations of every point on the sky, which in comparison to the reported number of detections revealed additional gaps in the astrometry and spectroscopy. \edits{Making these predictions required us to make the most precise, publicly-available determination of the \gaia scanning law. Using this scanning law, we further identified that most stars fainter than $G=22$ in DR2 have spurious magnitudes due to a miscalibration resulting from a thunderstorm over Madrid.} Our list of gaps \edits{and precision scanning law} will allow \edits{astronomers to know when \gaia's eye was truly on their binary star, exoplanet or microlensing event during the time period of the second data release}.
\end{abstract}




\section{Introduction}
The \gaia mission \citep{Gaia2016,Gaia2018} has revealed the dynamical richness of the Milky Way by composing unprecedentedly large astrometric, photometric and spectroscopic catalogues. It is important to remember, however, that we are viewing the Galaxy through the eyes of both \gaia and the \gaia Data Processing and Analysis Consortium (DPAC); \gaia is not equally sensitive to the properties of stars of every brightness and colour, whilst DPAC have applied stringent quality cuts to make the preliminary second data release (DR2) a useful resource. These biases in completeness are complex and are best described through a selection function, which gives the probability of a star making it into the published DR2 catalogue (or any subset that is of interest, such as the variable star or spectroscopic catalogues).

In this series of papers we will investigate the completeness of the \gaia data products by quantifying their selection functions. \edits{Apart from its predecessor \hipparcos \citep{hipShort,1997A&A...323L..49P},} the form of the \gaia selection functions are unlike any previous astronomical selection functions due to the spinning and precessing nature of the scanning law\edits{; there are two continuously roving fields-of-view rather than well-defined exposures of individual fields.} We intend to publish updates with each new \textit{Gaia} data release and for each catalogue of interest in each release.

\edits{The reason that the completeness of the \gaia DR2 catalogue depends on the scanning law is that the \gaia astrometric pipeline required that an object had been detected a minimum number of times in order for it to be included in \gaia DR2:}
\begin{itemize}
    \item \citet{Lindegren2018} required at least five \textsc{astrometric}\_\textsc{matched}\_\textsc{observations} \edits{(transits across the field of view that were used in the astrometric pipeline)} for a source to have a two-parameter astrometric solution and at least six \textsc{visibility}\_\textsc{periods}\_\textsc{used} for a five-parameter astrometric solution (a visibility period is a cluster of astrometric matched observations \edits{separated from other clusters by a gap of at least 4 days}, and thus \textsc{visibility}\_\textsc{periods}\_\textsc{used} $\leq$ \textsc{astrometric}\_\textsc{matched}\_\textsc{observations}). Note that a two-parameter astrometric solution is a minimum requirement for a source to be included in \gaia DR2. 
    \item \edits{The \gaia DR2 documentation\footnote{\label{note1}\url{https://gea.esac.esa.int/archive/documentation/GDR2/Catalogue_consolidation/chap_cu9cva/sec_cu9cva_consolidation/ssec_cu9cva_consolidation_ingestion.html}} mentions a further cut that the mean photometric $G$ measurement must have been derived from a minimum number of 10~\textsc{phot\_g\_n\_obs} ($G$-band CCD observations) for a source to have a two-parameter astrometric solution. This cut is much weaker than the $\textsc{astrometric}\_\textsc{matched}\_\textsc{observations}\geq5$ cut because each individual transit across the field of view can result in as many as nine \textsc{phot}\_\textsc{g}\_\textsc{n}\_\textsc{obs} due to the multiple columns of CCDs. This cut will mostly act to remove spurious sources, but may also cut some of the faintest sources.}
\end{itemize}
\edits{There were 1,692,919,135 sources that were included in DR2, but they may still be missing the other photometric and spectroscopic data products (see Table~1 of \citealp{Gaia2018} for a more comprehensive overview of the fractions of sources which were published with each kind of measurement). We list here the selection criteria of several of the most important \gaia subsets:}   
\begin{itemize}
    \item \citet{Riello2018} required at least two \textsc{phot}\_\textsc{bp}\_\textsc{n}\_\textsc{obs} or two \textsc{phot}\_\textsc{rp}\_\textsc{n}\_\textsc{obs} for a source to have blue or red photometry respectively.
    \item \citet{Sartoretti2018} required at least two \textsc{rv}\_\textsc{nb}\_\textsc{transits} for a source to have a spectroscopic radial velocity. 
    \item \edits{Photometric time series are only available in DR2 for the 550,737 
    sources classified as variable (\textsc{phot\_variable\_flag} set to `VARIABLE'). \cite{Holl2018} gives an overview of the variability pipeline (including cuts on a minimum number of filtered FoV-transits ranging from 5 to 20 depending on the variability product).}
\end{itemize}

Furthermore, such requirements are popular quality cuts, with \citet{Arenou2018} suggesting \textsc{visibility}\_\textsc{periods}\_\textsc{used} $>$ 8 and \citet{Marchetti2019} recommending \textsc{rv}\_\textsc{nb}\_\textsc{transits} > 5. These cuts \edits{are used to} select stars which have usable astrometry, photometry and spectroscopy and thus define the \edits{practical selection function} of \gaia DR2. \edits{Note that we will not discuss the Solar System Objects \citep[see ][and references therein]{2018A&A...616A..13G} as their processing is rather different from the above mentioned data products.}

A source can only have had a certain number of astrometric, photometric or spectroscopic detections if \gaia observed that source at least that many times. In practise, \gaia will need to have observed a source more than the minimum number of times, because not every observation results in a \edits{reported} detection (e.g. faint sources may not trigger a detection on every observation due to photon shot noise\edits{, and during Galactic plane scans a fraction of the detections can be deleted on-board due to a lack of bandwidth to down-link all of the data}). We can model this as a probabilistic process (that may depend on the property of the star and its environs), and then predict the selection function as the probability that enough of the observations resulted in detections for the source to clear the minimum threshold for inclusion in \gaia DR2. However, there are systematic reasons that a source might have fewer detections than expected: \edits{transmission loss due to contamination}, focal plane decontamination \edits{events to mitigate the former}, \edits{(sufficiently strong)} micro-meteoroid impacts\edits{, refocussing events} and station-keeping maneuvers all cause breaks in \gaia's scientific data-taking \citep{Gaia2016}. The \gaia pipelines also differ in whether they include data taken during the Ecliptic Pole Scanning Law (EPSL), where for the first month of observations \gaia scanned through the North and South Ecliptic Poles on every revolution (the photometric and spectroscopic pipelines include that data, the astrometric pipeline does not). These \edits{filters on the data} cause stars to lose observations (i.e. opportunities for detections) that they would otherwise have been granted by the nominal scanning law, and in a way that is systematic across the sky. If \gaia is not \edits{making useful detections} for six hours then all sources within a  $0.7^{\circ}$-wide strip around the sky will have lost one nominal observation. By neglecting these gaps we would be introducing a systematic spatial-bias into our selection function.

The objective of this paper is thus to work out the times at which \gaia was obtaining the data that was used in each of the astrometric, photometric and spectroscopic pipelines. \edits{We will refer to the time periods during Gaia DR2 data-taking that did not contribute any data to the DR2 data products as \emph{gaps}, and the fraction of observations made outside of these gaps that resulted in detections used in the DR2 data products as the \emph{efficiency}.} \edits{Along the way we will create the most precise publicly-available determination of the \gaia scanning law. These are fundamental ingredients} that we will use in the later papers of this series to deduce the \gaia selection functions.

\section{Methodology}
\label{sec:methodology}

\gaia DR2 covers the time period of 2014 July 25 (10:30 UTC) until 2016 May 23 (11:35 UTC). A more natural clock used by the \gaia DPAC is the on-board mission timeline (OBMT), which counts the nominal number of revolutions that \gaia has completed. \gaia completes four revolution per day and so we can express the OBMT in either revolutions or days. An approximate relation between OBMT (in revolutions) and barycentric coordinate time (TCB, in Julian years) at \gaia is given by \citet{Lindegren2018}:
\begin{equation}
    \mathrm{TCB} \simeq \mathrm{J2015.0} + (\mathrm{OBMT}-1717.6256\; \mathrm{rev})/1461\;\mathrm{rev}\;\mathrm{yr}^{-1}.
\end{equation}
The beginning of \gaia scientific data-taking occurs at $1078.3795\;\mathrm{rev}$ and the final data included in \gaia DR2 was taken at $3750.5602\;\mathrm{rev}$. We will use the OBMT time-system throughout the rest of this paper, showing the OBMT time wherever possible in both days and revolutions.

\gaia followed the Ecliptic Pole Scanning Law (EPSL) for the first month of science operations until  $1192.13\;\mathrm{rev}$ (2014 August 22 21:00 UTC). This period was excluded from the astrometric pipeline, but was retained for the photometric and spectroscopic pipelines. There were two other significant data-taking breaks which affected all of the pipelines \citep{Gaia2016}: the decontamination events on 23 September 2014 ($\mathrm{OBMT} \approx 1317\;\mathrm{rev}$) and on 3 June 2015 ($\mathrm{OBMT} \approx 2330\;\mathrm{rev}$). During these two periods the focal planes and some mirrors were heated to sublimate the water ice that had condensed on their surfaces and which was degrading their optical properties. These events resulted in days of lost data-taking opportunity, because \gaia could not resume taking useful scientific measurements until the focal plane had returned to its nominal operating temperature \citep{Riello2018}. Following each of these decontaminations, refocussing of the optics occurred at $\mathrm{OBMT} \approx 1443.950\;\mathrm{rev}$ and $\mathrm{OBMT} \approx 2574.644\;\mathrm{rev}$, however these gaps are relatively minor as they lasted for less than one hour. The \gaia DPAC has reported many other potential gaps in \gaia data-taking\footnote{\url{https://gea.esac.esa.int/archive/documentation/GDR2/Introduction/chap_cu0int/cu0int_sec_release_framework/cu0int_sssec_spacecraft_status.html}}, but have only published rough times rather than precise intervals.

In the remainder of this section we will use the epoch photometry released as part of \gaia DR2 to identify many of these uncertain gaps, and then address in turn the further technical gaps that are relevant for the astrometric, photometric and spectroscopic pipelines.

\subsection{Misusing the epoch photometry of variable stars}
\label{sec:epoch}

\begin{figure}
	\centering
	\includegraphics[width=1.\linewidth,trim=0 0 0 0, clip]{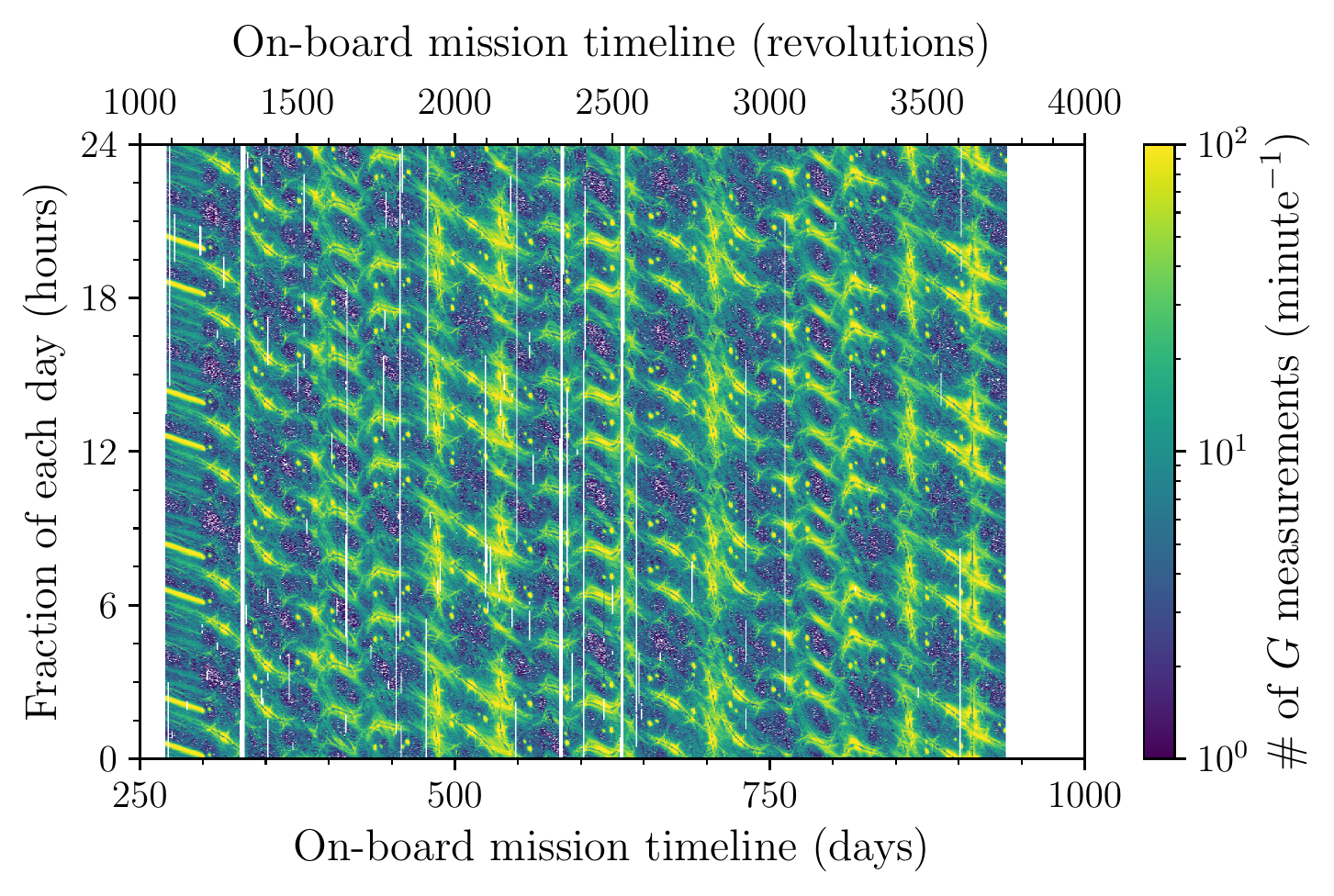}
	a) A map of the number of $G$ measurements made per minute.\vspace{0.5cm}
	
	\includegraphics[width=1.\linewidth,trim=0 0 0 0, clip]{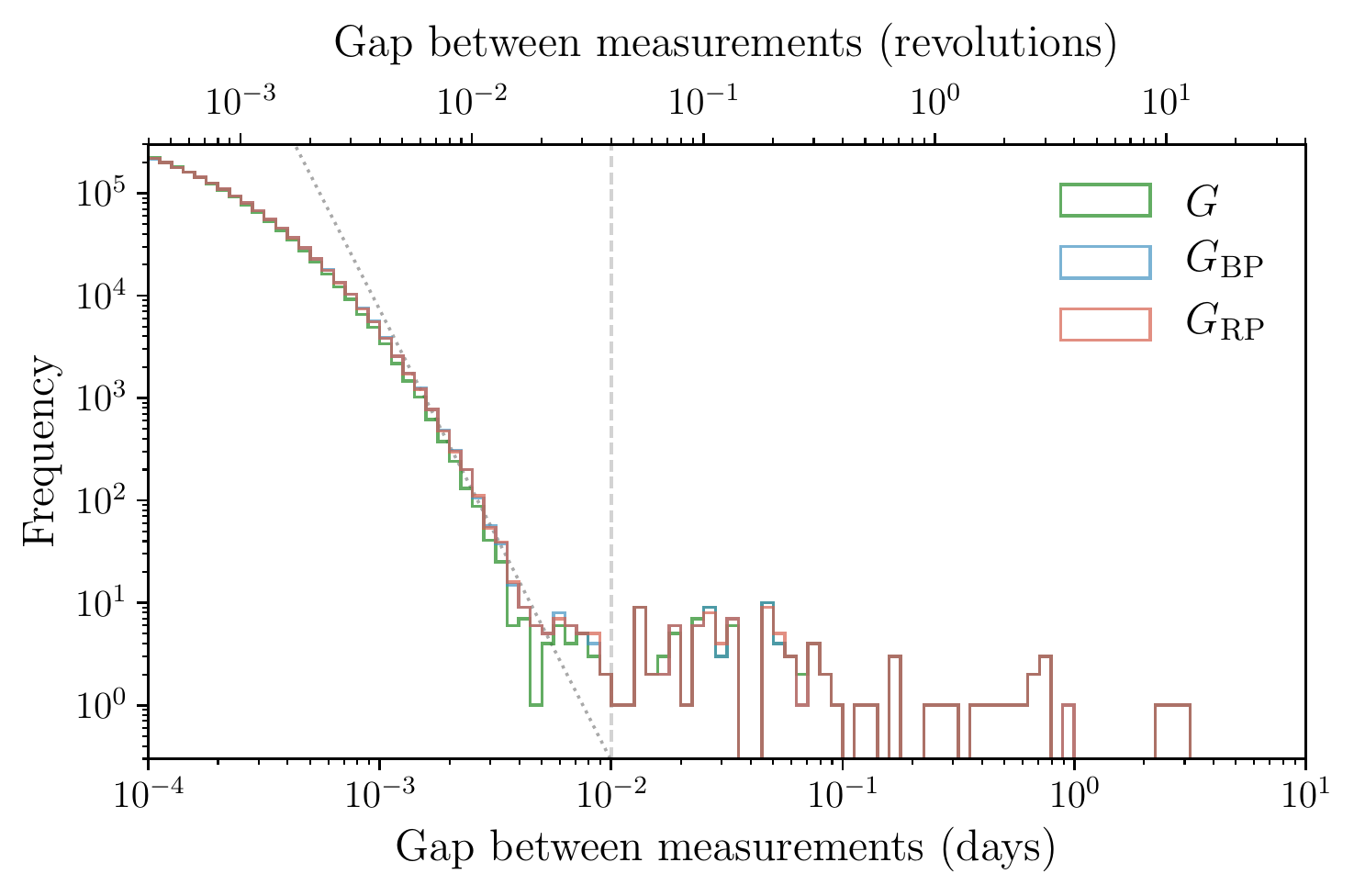}
	b) Difference between consecutive epoch photometric measurements.\vspace{0.5cm}
	
	\includegraphics[width=1.\linewidth,trim=0 0 0 0, clip]{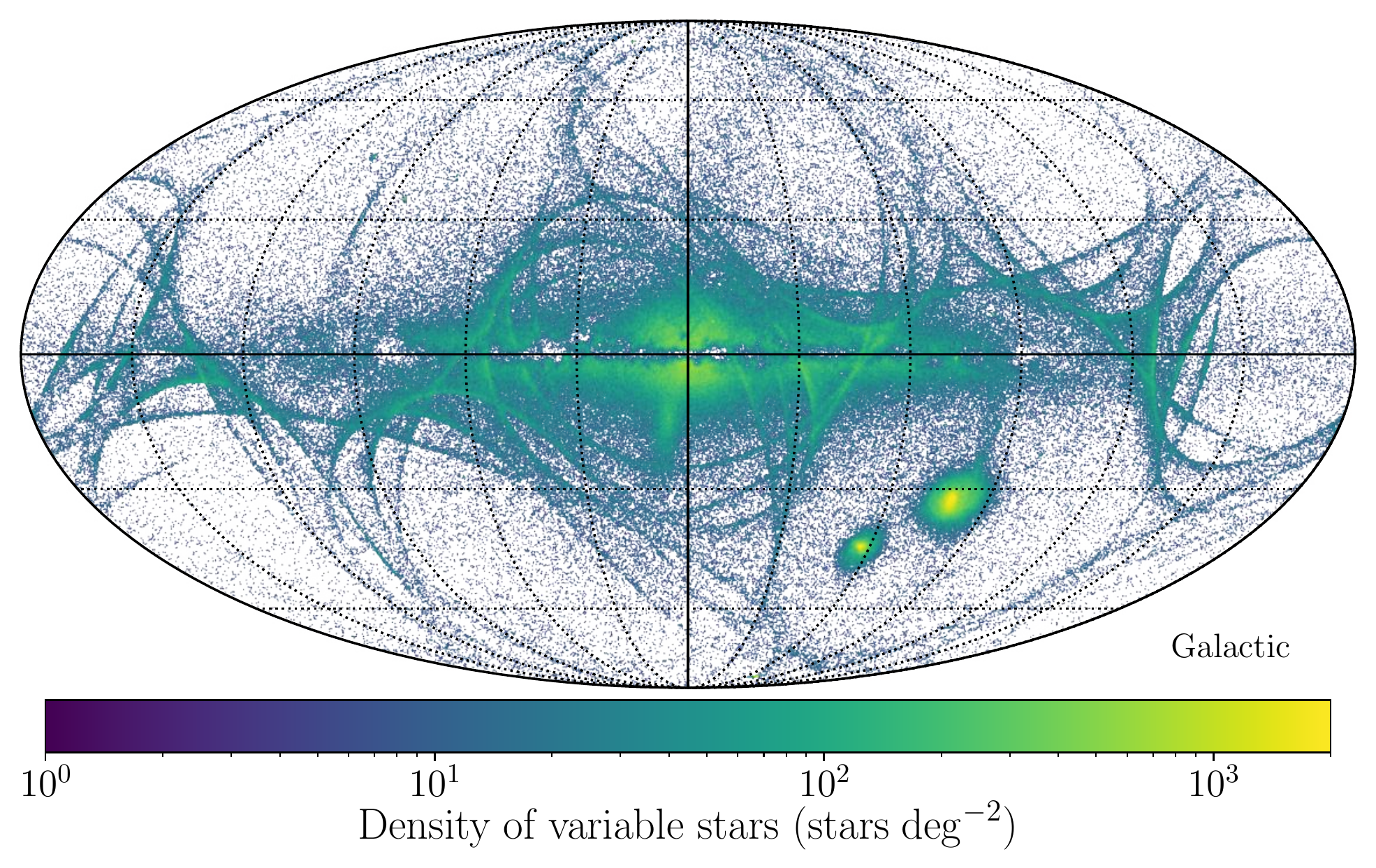}
	c) Location and density of stars with \gaia DR2 epoch photometry.
	\caption{The epoch photometry of 550,737 variable stars was available in \gaia DR2 and we used the times of these measurements to detect 89 gaps in the \gaia data-taking. \textbf{Top:} A map of the number of $G$ epoch measurements of these 550,737 stars per minute of the \gaia DR2 period. Much of the variation is due to the varying stellar density at different locations on the sky, but gaps in the \gaia data-taking are visible as white vertical streaks. This panel was inspired by Fig. 2 of \citet{Riello2018}. \textbf{Middle:} A log-log histogram of the time difference between consecutive epoch photometric measurements. Most of the differences fall in a distribution with a tail indicated by the dotted line and that ends by the dashed line at $10^{-2}\;\mathrm{days}$. The 89 differences longer than this are due to genuine breaks in the \gaia data-taking. \textbf{Bottom:} A map of the sky in Galactic coordinates where each point corresponds to one of the 550,737 variable stars. The colour of each point indicates the local density of these stars.}
	\label{fig:epoch}
\end{figure}

\begin{figure*}
	\centering
	\includegraphics[width=1.\linewidth,trim=70 110 35 110, clip]{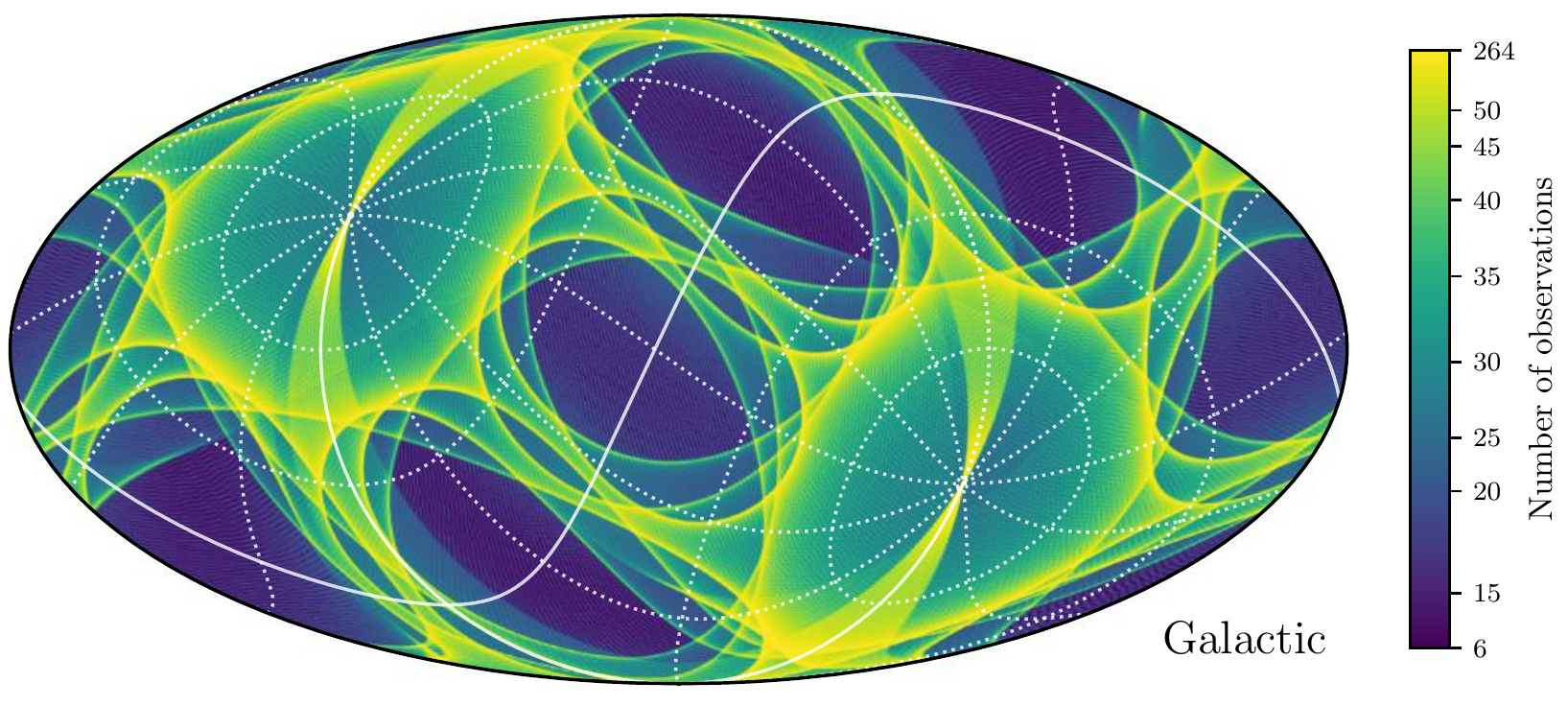}
	\caption{A \edits{Galactic} HEALPix map of the estimated number of times that \gaia looked at each location on the sky \edits{during the DR2 time range, as calculated using our determination of the scanning law without accounting for data-taking gaps}. The graticule marks out lines of longitude and latitude in the Ecliptic coordinate frame, which is the native frame of the \gaia scanning law. The colourmap has been histogram-equalised to increase the contrast.}
	\label{fig:map}
\end{figure*}

\gaia is not only measuring the astrometric, photometric and spectroscopic properties of billions of stars across the entire sky, it is measuring each of those properties tens or hundreds of times across a five-to-ten year period. The quantities given in \gaia DR2 were simply summary statistics of these epoch measurements. However, \gaia DR2 did contain epoch broadband photometry for the 550,737 sources that had been classified as variable \edits{\citep[see][and references therein]{Holl2018}}. If there was a gap in data-taking then there could be no epoch photometric measurements taken during that gap to report, and thus gaps in data-taking will appear as periods where none of these variable stars have reported measurements. We illustrate this principle in the top panel of Fig. \ref{fig:epoch}, where we show the number of epoch $G$ band measurements of this set of variable stars during each minute of \gaia DR2. We stress that this figure was inspired by Fig. 2 of \citet{Riello2018}. The plot is arranged so that each column is one day of the mission and the rows give the measurement frequency at each minute of each day. This can be considered to be a projection of the source density that each field-of-view is observing through time, with that pattern repeating four times per day due to \gaia's rotation and evolving over longer timescales due to \gaia's precession. The EPSL appears on the left-hand side as a smearing of the more intricate pattern seen during the nominal scanning law. Gaps in \gaia's photometric data-taking are clearly visible as white streaks in this figure. \edits{It is important to note that the times reported in the \gaia DR2 Epoch Photometry table have been corrected to the Solar system barycentre, while we are interested in the time on-board \gaia. Rather than attempt that correction ourselves, we note that the \textsc{transit}\_\textsc{id} column of the DR2 epoch photometry table encodes the field-of-view, CCD and pixel of each transit, in addition to the OBMT time that the source was detected on the first column of the astrometric CCDs (AF1). We use that time in place of the transit or observation times. We have ignored all of the rejection flags provided with the epoch photometry which identified detections flagged by the photometric and variability pipelines and so maximised the information available to us on the observation epochs, because bad detections still have valid observation times.}

To identify the intervals of the gaps we ordered the epoch photometry by time for each of the bands and calculated the time difference between consecutive measurements. We show a log-log histogram of these time differences in the middle panel of Fig. \ref{fig:epoch}. If there were not data-taking gaps then we would still expect there to be a distribution of time differences because of the varying angular distances between consecutively scanned sources (c.f. the bottom panel of Fig. \ref{fig:epoch} where we show the sky distribution of the variable stars). Almost all of the differences appear to lie in a distribution that falls off as a power law, suggesting that we shouldn't find any differences longer than $1\%$ of a day (864 seconds or 4\% of a revolution). Contrary to this expectation, we find that there are \edits{94} gaps in the $G$ photometry that are longer than this, with the longest lasting for $3.11\;\mathrm{days}$ beginning at $\mathrm{OBMT}\approx 1316.37$ and thus corresponding to the first mirror decontamination. There are also \edits{94} gaps in each of the $G_{\mathrm{BP}}$ and $G_{\mathrm{RP}}$ epoch photometry series, which correspond to the same \edits{94} gaps that are in the $G$ epoch photometry series. In all but one case these gaps are of approximately the same length, but the gap that begins at $\mathrm{OBMT}\approx2751.4\;\mathrm{rev}$ (approximately at the end of the time period included in \gaia DR1) is only \edits{$0.27\;\mathrm{rev}$} long in the $G$ series but \edits{$3.66\;\mathrm{rev}$} in the $G_{\mathrm{BP}}$ and $G_{\mathrm{RP}}$ series. To make our estimation of the gaps as accurate as possible, we merged the time series from the three bands together and re-identified the \edits{94} gaps longer than $1\%$ of a day. We then combined the $G_{\mathrm{BP}}$ and $G_{\mathrm{RP}}$ series together to identify that the gap at the end of \gaia DR1 has an extension \edits{$\mathrm{OBMT} = 2751.605 {-} 2755.000\;\mathrm{rev}$} in the colour photometry. \edits{We note that we do not recover the first of the two refocussing events due its short length (0.625\% of a day) but that we do recover the second}.

\edits{There are only three gaps longer than one day: the two gaps associated with the decontaminations and a third gap beginning at \edits{$\mathrm{OBMT}\approx 2524.478\;\mathrm{rev}$} and lasting for \edits{$9.442\;\mathrm{rev}$} that does not appear to be listed in the DPAC literature or documentation. This gap (\gaia Helpdesk, private communication) can be traced back to an issue in the software used to decompress the downlinked data that prevented DPAC from generating the basic reconstructed star packet for the astrometric (and photometric) processing. This gap will be closed in a future data release because the raw data has since been re-processed and so will enter the processing in later releases.}

We list all \edits{94} of these gaps in Tab. \ref{tab:gapsepoch}. We note that these gaps are only approximate. We have assumed that the time that a gap starts and ends is the time of the last observation before the gap and the first observation after the gap, respectively, which is merely a good approximation. \edits{Additionally, the times we have used are the times of acquisition on the first astrometric CCD (AF1), and thus the actual time at which the observation was taken on the $G$, $G_{\mathrm{BP}}$ or $G_{\mathrm{RP}}$ CCDs could differ by as much as 40 seconds (0.05\% of a day).}

\begin{table}
\caption{Overview of the gaps in \gaia data-taking that were identified from the epoch photometry and which apply to all \gaia DR2 data products. All times are given in units of revolutions. \edits{This table is available electronically at \url{https://doi.org/10.7910/DVN/P8FVOZ}, excluding the additional gap for $G_{\mathrm{BP}}$ and $G_{\mathrm{RP}}$.}}
\label{tab:gapsepoch}
\setlength{\cmidrulekern}{0.1em}
\setlength\tabcolsep{4.5pt}
\begin{tabular}{llr|llr}
\hline \multicolumn{6}{c|}{Gaps from epoch photometry that apply to all of \gaia DR2} \\ 
\hline Start & End  & Length & Start & End  & Length \\ \cmidrule(l{4pt}r{4pt}){1-3} \cmidrule(l{4pt}r{4pt}){4-6}
1088.119  &  1090.200  &   2.081  &  1820.689  &  1823.857  &   3.168  \\
1100.441  &  1100.493  &   0.052  &  1828.059  &  1828.197  &   0.138  \\
1105.081  &  1105.399  &   0.318  &  1828.405  &  1828.945  &   0.540  \\
1148.242  &  1148.307  &   0.066  &  1829.075  &  1829.119  &   0.043  \\
1185.160  &  1185.354  &   0.194  &  1907.711  &  1910.519  &   2.809  \\
1189.162  &  1189.355  &   0.192  &  1919.017  &  1919.126  &   0.109  \\
1212.046  &  1212.112  &   0.066  &  1943.453  &  1943.554  &   0.101  \\
1241.838  &  1241.893  &   0.055  &  1951.340  &  1951.535  &   0.195  \\
1243.861  &  1243.917  &   0.056  &  1962.368  &  1962.480  &   0.112  \\
1261.361  &  1261.568  &   0.207  &  2093.999  &  2095.570  &   1.571  \\
1297.890  &  1297.939  &   0.049  &  2099.221  &  2099.414  &   0.193  \\
1311.211  &  1311.290  &   0.079  &  2103.599  &  2103.675  &   0.076  \\
1312.033  &  1312.085  &   0.052  &  2111.238  &  2111.485  &   0.248  \\
1316.060  &  1316.113  &   0.053  &  2139.415  &  2139.537  &   0.122  \\
1316.365  &  1328.800  &  12.435  &  2142.287  &  2142.396  &   0.109  \\
1335.617  &  1335.748  &   0.132  &  2150.428  &  2150.541  &   0.112  \\
1336.675  &  1336.788  &   0.113  &  2154.118  &  2154.226  &   0.107  \\
1380.713  &  1380.900  &   0.187  &  2172.833  &  2173.071  &   0.239  \\
1384.163  &  1384.271  &   0.108  &  2179.634  &  2179.765  &   0.131  \\
1388.222  &  1388.275  &   0.054  &  2192.250  &  2195.219  &   2.969  \\
1401.750  &  1402.066  &   0.316  &  2233.849  &  2233.922  &   0.073  \\
1403.511  &  1403.609  &   0.098  &  2233.934  &  2234.022  &   0.088  \\
1404.059  &  1404.114  &   0.055  &  2235.656  &  2235.854  &   0.198  \\
1404.362  &  1404.703  &   0.340  &  2246.646  &  2246.842  &   0.196  \\
1471.939  &  1472.240  &   0.301  &  2330.539  &  2341.480  &  10.942  \\
1484.512  &  1484.565  &   0.052  &  2354.221  &  2355.449  &   1.228  \\
1498.047  &  1498.243  &   0.197  &  2371.934  &  2372.249  &   0.316  \\
1498.300  &  1498.373  &   0.072  &  2405.965  &  2408.645  &   2.680  \\
1516.821  &  1517.207  &   0.386  &  2408.932  &  2409.969  &   1.037  \\
1517.402  &  1517.498  &   0.097  &  2472.230  &  2472.369  &   0.138  \\
1517.596  &  1517.691  &   0.095  &  2499.490  &  2499.682  &   0.192  \\
1517.790  &  1517.885  &   0.095  &  2524.478  &  2533.920  &   9.442  \\
1517.988  &  1518.085  &   0.098  &  2574.644  &  2576.320  &   1.676  \\
1527.066  &  1527.166  &   0.100  &  2576.329  &  2576.545  &   0.216  \\
1606.501  &  1606.701  &   0.200  &  2583.929  &  2584.265  &   0.336  \\
1623.571  &  1623.696  &   0.125  &  2592.295  &  2592.365  &   0.071  \\
1651.166  &  1651.836  &   0.670  &  2651.926  &  2651.988  &   0.062  \\
1652.333  &  1652.456  &   0.122  &  2751.337  &  2751.605  &   0.268  \\
1653.557  &  1656.025  &   2.468  &  2922.026  &  2922.697  &   0.671  \\
1731.875  &  1731.977  &   0.101  &  2922.755  &  2923.406  &   0.651  \\
1770.003  &  1770.499  &   0.496  &  2923.718  &  2923.995  &   0.277  \\
1773.708  &  1773.837  &   0.129  &  2924.291  &  2924.431  &   0.139  \\
1776.937  &  1777.175  &   0.238  &  3045.105  &  3048.186  &   3.080  \\
1788.113  &  1788.173  &   0.060  &  3205.136  &  3205.188  &   0.052  \\
1811.568  &  1812.469  &   0.901  &  3254.085  &  3254.288  &   0.204  \\
\hline \multicolumn{3}{c|}{Additional gap for $G_\mathrm{BP}$/$G_\mathrm{RP}$} & 2751.605 & 2755.000 & 3.395 \\ 
\hline
\end{tabular}
\end{table}

\subsection{Predicting the number of observations}
\label{sec:predicting}

Our second approach to identifying gaps in \gaia's scientific data-taking was to predict the number of times that each \edits{of the 1,692,919,135 sources} was observed during \gaia DR2 and compare to the observed number of detections. We define an observation to be a source entering into one of the \gaia fields-of-view (FoV). A key feature of \gaia's design is that it has two fields-of-view separated by a basic angle of $106.5^{\circ}$, which allows \gaia to perform absolute astrometry because the light from the two FoVs is combined to overlap onto the single \gaia focal plane. Due to technical considerations the centres of the two fields-of-view do not lie at the centre of the focal plane, but are instead offset by approximately $221\;\mathrm{arcsec}$ \edits{in the across-scan direction (i.e. the direction orthogonal to the scanning direction, as opposed to the along-scan direction which is parallel to the scanning direction)}. The number of times that a source crosses the FoVs is determined by the spinning-and-precessing scanning law. \gaia spins with a period of 6 hours in the plane of the two FoVs, such that each source that crosses the preceding field-of-view typically crosses the following field-of-view less than two hours later. The absolute direction of the spin axis is fixed to be $45^{\circ}$ away from the direction to the Sun, and the spin axis precesses about this direction with a period of 63 days. This precession causes sources to cross the FoVs at a slight angle and thus it is possible for a source to cross the preceding FoV but not the following FoV, and vice-versa.

\edits{The number of times $n$ that each source should have crossed the field of view according to the scanning law (hereafter referred to as an observation\footnote{Note that the column \textsc{matched\_observations} gives the number of detections of the source by \gaia, which each occurred on one of the possible observations predicted by the scanning law.}) was not provided in the \gaia DR2 catalogue, however the \gaia DPAC has recently published\footnote{\url{https://www.cosmos.esa.int/web/gaia/scanning-law-pointings}} the nominal scanning law which gives the location of the centres of the preceding and following fields-of-view at $10\;\mathrm{sec}$ intervals for the 22 month period covering \gaia DR2, which in principle should allow us to estimate the number of observations ourselves. However, the true scanning law carried out by \gaia can have deviated by up to $30\;\mathrm{arcsec}$ from the commanded nominal one, which could cause us to predict observations of sources that did not occur or miss observations that did occur. In Appendix \ref{sec:pitchandroll} we use the information encoded in the \gaia DR2 epoch photometry table to determine the true \gaia scanning law to within $100\;\mathrm{mas}$. Our determination of the scanning law is publicly available as an accompaniment to this paper, and we will use our determination in the remainder of this work.}

\edits{To avoid computing the number of observations for every source in \gaia DR2, we instead computed the number of observations $n$ received by each pixel of an \textsc{nside}=4096, Equatorial, nested HEALPix \citep{Gorski2005} map, and then approximated the number of observations of each source by bilinearly interpolating the number of observations of the nearest eight pixel centres and then rounding to the nearest integer. This is likely to be a reasonable approximation because the pixels in this map are roughly \edits{$0.014\;\mathrm{deg}$} in size, which is significantly smaller than the $0.7^{\circ}$ size of the \gaia FoVs that set the angular scale of the scanning law. To verify the accuracy of this approximation we generated 1,000,000 points uniformly distributed across the sky, computed their exact number of observations according to our scanning law, and then estimated the number of observations using our interpolated approximation. The differences between the true and estimated number of observations had a standard deviation of $0.504$ and only 0.76\% percent fell outside the interval $[-1,+1]$, with the difference being zero in 76.95\% of cases. We determined that the error distribution had a heavier-than-Gaussian tail and was better approximated by a symmetric Skellam distribution\footnote{The distribution of the difference between two independent and identically distributed Poisson random variables.} of the same variance.}

\edits{We stepped through the timeseries of FoV centres and at each timestep logged all of the HEALPix pixel centres that were within $0.7^{\circ}$ of either of the FoV centres and whose location was consistent with falling onto the CCDs in the focal plane (in the case of the spectroscopic measurements, we took into account that only four of the seven rows have radial velocity spectrometer CCDs). In actual fact, the star needs to cross over the column of Skymapper CCDs for it to be assigned a window, and due to across-scan motion it is possible for stars to enter the \gaia FoV after the Skymapper column. However, given the small size of the across-scan motion and the impact of this motion relative to other uncertainties in our method, we opted to neglect this effect. After discarding duplicate observations (pixel centres take 42 seconds to cross the FoV and so are logged multiple times by this procedure) and observations where the location of the pixel centre on the focal plane was outside the CCD array (neglecting the small gaps between CCD rows, as done by the official DPAC Gaia Observation Forecasting Tool, \url{https://gaia.esac.esa.int/gost/}), the number of logged observations is an estimate of the number of observations for each pixel centre. The overlaps between the precessing scans of \gaia produces an intricate pattern in the number of observations across the sky with angular structure at all scales, which we illustrate in Fig. \ref{fig:map}.}

\subsection{Comparing predictions to observations}
\label{sec:comparison}

\begin{table}
\caption{Overview of the gaps in \gaia data-taking that were identified from the \gaia DPAC papers or through investigation in this work.}
\label{tab:gapstechnical}
\begin{tabular}{llrllr}
\hline Start     & End       & Length & \multicolumn{3}{c|}{Explanation} \\
\hline \multicolumn{6}{c|}{Additional technical gaps in the astrometry} \\ \hline 
1078.378 & 1192.130 & 113.751 & \multicolumn{3}{c|}{Ecliptic Pole Scanning Law} \\
1316.490 & 1389.113 & 72.623 & \multicolumn{3}{c|}{Decontamination} \\
2179.125 & 2191.000 & 11.875 & \multicolumn{3}{c|}{Meteoroid hit during patch} \\
2324.900 & 2401.559 & 76.659 & \multicolumn{3}{c|}{Decontamination} \\
\hline \multicolumn{6}{c|}{Additional technical gaps in the photometry} \\ \hline 
1078.378 & 1081.000 & 2.621 & \multicolumn{3}{c|}{Ecliptic Pole Scanning Law} \\
1316.492 & 1324.101 & 7.609 & \multicolumn{3}{c|}{Decontamination} \\
2330.616 & 2338.962 & 8.346 & \multicolumn{3}{c|}{Decontamination} \\
\hline \multicolumn{6}{c|}{Additional technical gaps in the spectroscopy} \\ \hline 
1316.492 & 1360.000 & 43.508 & \multicolumn{3}{c|}{Decontamination} \\
2183.100 & 2226.800 & 43.700 & \multicolumn{3}{c|}{Meteoroid hit during patch} \\
2324.900 & 2384.000 & 59.010 & \multicolumn{3}{c|}{Decontamination} \\ 
\hline \multicolumn{6}{c|}{Additional technical gaps that apply to all of \gaia DR2} \\ \hline 
1443.950 & 1443.975 & 0.025 & \multicolumn{3}{c|}{Refocussing} \\
2574.644 & 2574.728 & 0.084 & \multicolumn{3}{c|}{Refocussing} \\ \hline
\end{tabular}
\end{table}

If \gaia successfully detected a source astrometrically, photometrically and spectroscopically on every observation, then testing our prediction would be straightforward; we could simply map the difference between the number of \edits{astrometric, photometric or spectroscopic} detections $k$ \edits{reported in the \gaia DR2 table} and the \edits{number of observations $n$ that we predict using the scanning law}. However, we know that \gaia is not perfectly efficient. Not every observation results in a detection used in \gaia DR2. This is particularly true for faint sources in crowded regions; the effective crowding limit for the astrometry and $G$ photometry is $1,050,000\;\mathrm{sources}\;\mathrm{deg}^{-2}$ \citep{Gaia2016}, beyond which faint stars begin to miss out on detections. Despite this, we should expect that the detection efficiency for stars brighter than $G\lesssim16$ should be near 100\%, and thus \edits{that} the maximum number of detections $k$ of any source in each pixel of an \textsc{nside}=128 HEALPix map should be approximately \edits{equal to the number of observations of that pixel}. Any significant differences between $k$ and the predicted number of observations $n$ should thus be due to \edits{periods of time when \gaia data-taking was not resulting in useful detections}. To compare apples-to-apples, we downsampled our \edits{\textsc{nside}=4096} HEALPix map of predicted observations by finding the maximum of the \edits{1024} sub-pixels that make up each \textsc{nside}=128 pixel. We show maps of the difference $k-n$ in each of the following subsections.

This approach fails when searching for gaps in the spectroscopic data-taking. After removing all of the gaps identified in the remainder of this paper, the median of the ratio of the maximum number of detections in a pixel to predicted number of observations for the astrometry and colour photometry is 100\% and \edits{95.2\%} respectively, while for the spectroscopy it is still only \edits{66.6\%}. \edits{This is likely due to extensive cuts applied in the RVS pipeline that removed any detections where the window overlapped with a window assigned to another source (see \citealp{Sartoretti2018} for more details or Sec.~3.2 of \citealp{Boubert2019} for a condensed description). An improved algorithm that can handle overlapping windows is being written in advance of \gaia DR3 and so this effective efficiency will improve in future data releases. However, in \gaia DR2, }the maximum number of radial velocity detections is not at all a good estimator for the true number of radial velocity spectrometer observations. Our solution was \edits{adopt a second methodology where we} unfold the \textsc{nside}=128 HEALPix map of the deviancy $D\equiv(k-n)/n$ as a function of time. We did this by taking the ten second interval timeseries of the FoV centres and at each time taking the value of $D$ of the nearest pixel to each FoV. We then added these two time series together element-wise to emphasise the signal of the gaps. The motivation for doing this is that gaps in the data-taking will cause the deviance $D$ to grow more negative, and by aligning these deviances in time the gaps in data-taking should be clear. To make the signal visible to the naked eye, we had to apply a first-order Savitzky-Golay filter with a 2161-step window (there are 2160 ten-second intervals in one OBMT revolution). We calculated these time series for the astrometry, photometry and spectroscopy, and show them in Fig. \ref{fig:deviancy}. Also shown in that plot are the gaps in the epoch photometry identified in Sec. \ref{sec:epoch}, the technical gaps in each pipeline which are discussed in the following sections, and text which describes some of the causes of the gaps.

\afterpage{%
\begin{landscape}
\begin{figure}
	\centering
	\includegraphics[width=0.95\linewidth,trim=75 36 50 55, clip]{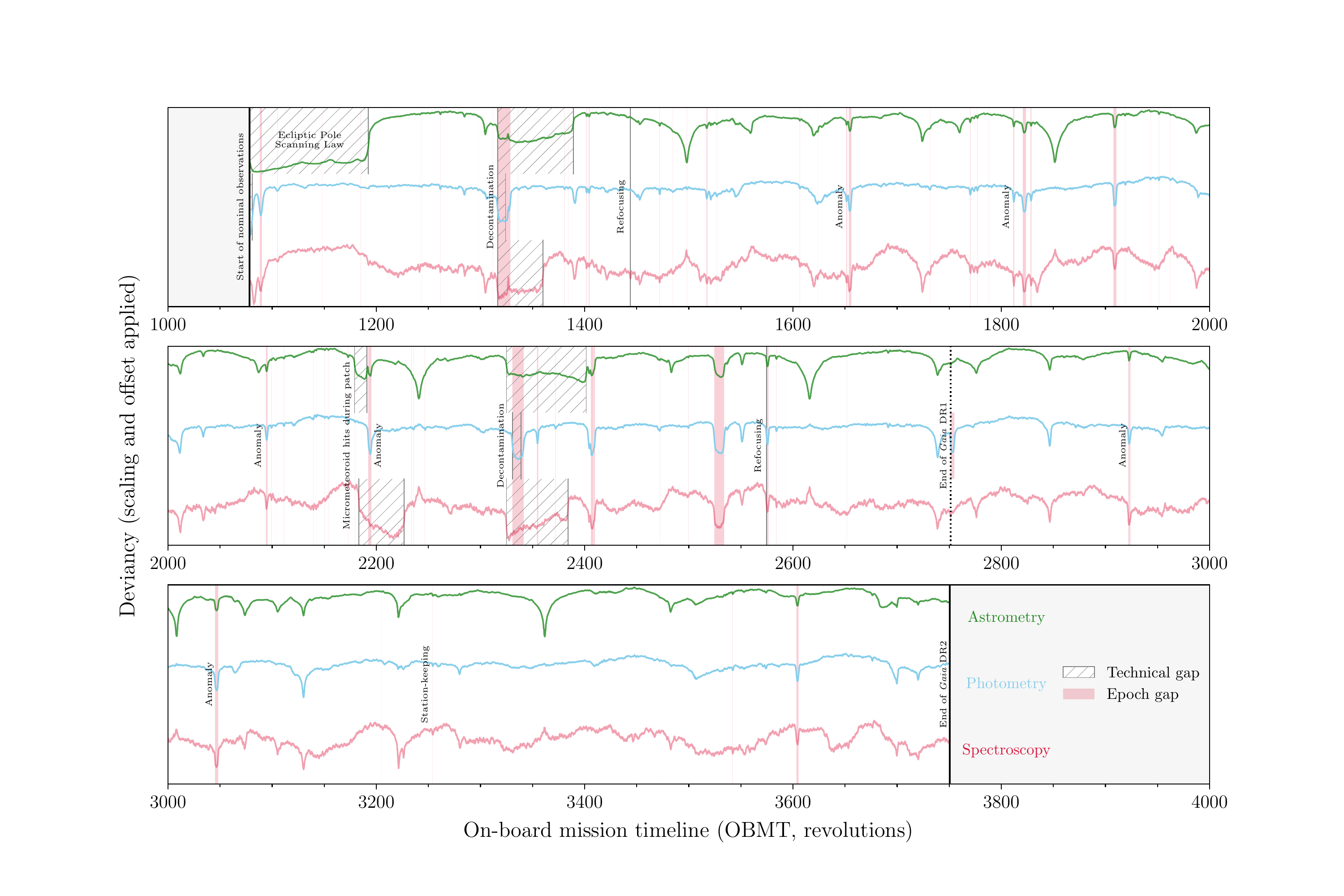}
	\caption{The gaps in the \gaia astrometry (green), colour photometry (blue) and spectroscopy (red) data-taking are visible as dips in these three deviancy time series. Deviancy is defined in Sec. \ref{sec:comparison} and is equal to the mean fractional difference between the maximum observed number of detections minus the predicted number of observations of stars in the pixels nearest to the two \gaia fields-of-view at each time. We have applied a smoothing filter with a one-revolution window to make the trend visible. We show the deviancy for the astrometry (\textsc{astrometric}\_\textsc{matched}\_\textsc{observations}), colour photometry (maximum of \textsc{phot}\_\textsc{bp}\_\textsc{n}\_\textsc{obs} and \textsc{phot}\_\textsc{rp}\_\textsc{n}\_\textsc{obs}) and spectroscopy (\textsc{rv}\_\textsc{nb}\_\textsc{transits}), and have applied scaling and offsets for visualisation purposes. The gaps identified from the \gaia DR2 epoch photometry and listed in Tab. \ref{tab:gapsepoch} are shown in red, while the gaps identified in the \gaia DPAC papers by this work and listed in Tab. \ref{tab:gapstechnical} are hatched. Textual annotations give explanations for some of the major gaps.}
	\label{fig:deviancy}
\end{figure}
\end{landscape}
}

We will now discuss each of the pipelines in turn, identifying specific technical gaps and their causes, and give all of those gaps in Tab. \ref{tab:gapstechnical} (we note that all of the pipelines had additional gaps due to refocussing, which are both given in that table). \edits{We do not apply our methodology to the number of \textsc{matched\_observations} (which gives the total number of detections assigned to the source) because this includes detections which are known to be spurious duplicates (as discussed in Sec. \ref{sec:duplicates}) which would violate the assumption of our methodology that the number of detections is less than or equal to the number of observations.}

\subsubsection{Astrometry}


The number of individual \gaia  \edits{ detections used in the astrometric pipeline} is given by the quantity \textsc{astrometric}\_\textsc{matched}\_\textsc{observations}, which in this section we will refer to as $k$. We show the difference between $k$ and $n$ in Fig. \ref{fig:astrometry}. The first panel is rich in structure but is dominated by the gaps mentioned by \citet{Lindegren2018}: the removal of the Ecliptic Pole Scanning Law (EPSL) at $\mathrm{OBMT}=1078.38{-}1192.13\;\mathrm{rev}$ and the decontaminations at $\mathrm{OBMT}=1316.490{-}1389.113\;\mathrm{rev}$ and $\mathrm{OBMT}=2324.900{-}2401.559\;\mathrm{rev}$. Accounting for these gaps results in the relatively clean second panel, and further removing the gaps we identified in the epoch photometry (astrometric measurements use the same CCDs as the $G$ photometry) gives the third panel. The only remaining clear gap is a previously undisclosed gap in the \gaia astrometric data-taking at $\mathrm{OBMT}=2179.125{-}2191.000\;\mathrm{rev}$, which was due to a micro-meteoroid hitting \gaia close to on-board software patch activities and resulting in a three-day-long deviation in the basic angle (\gaia Helpdesk, private communication). Removing this remaining gap results in the almost perfect fourth panel. The residual faint imprint of the \gaia scanning law at high latitudes is simply because at these locations there are so few stars that the maximum number of detections of these stars isn't quite equal to the true number of observations. There are perhaps additional gaps visible in the Southern hemisphere, but as we do not know whether these are true gaps or just drops in detection efficiency (see the following section for an example of this) we opted to leave them in.

\subsubsection{Photometry}

\begin{figure*}
	\centering
	\includegraphics[width=0.85\linewidth,trim=10 105 0 5, clip]{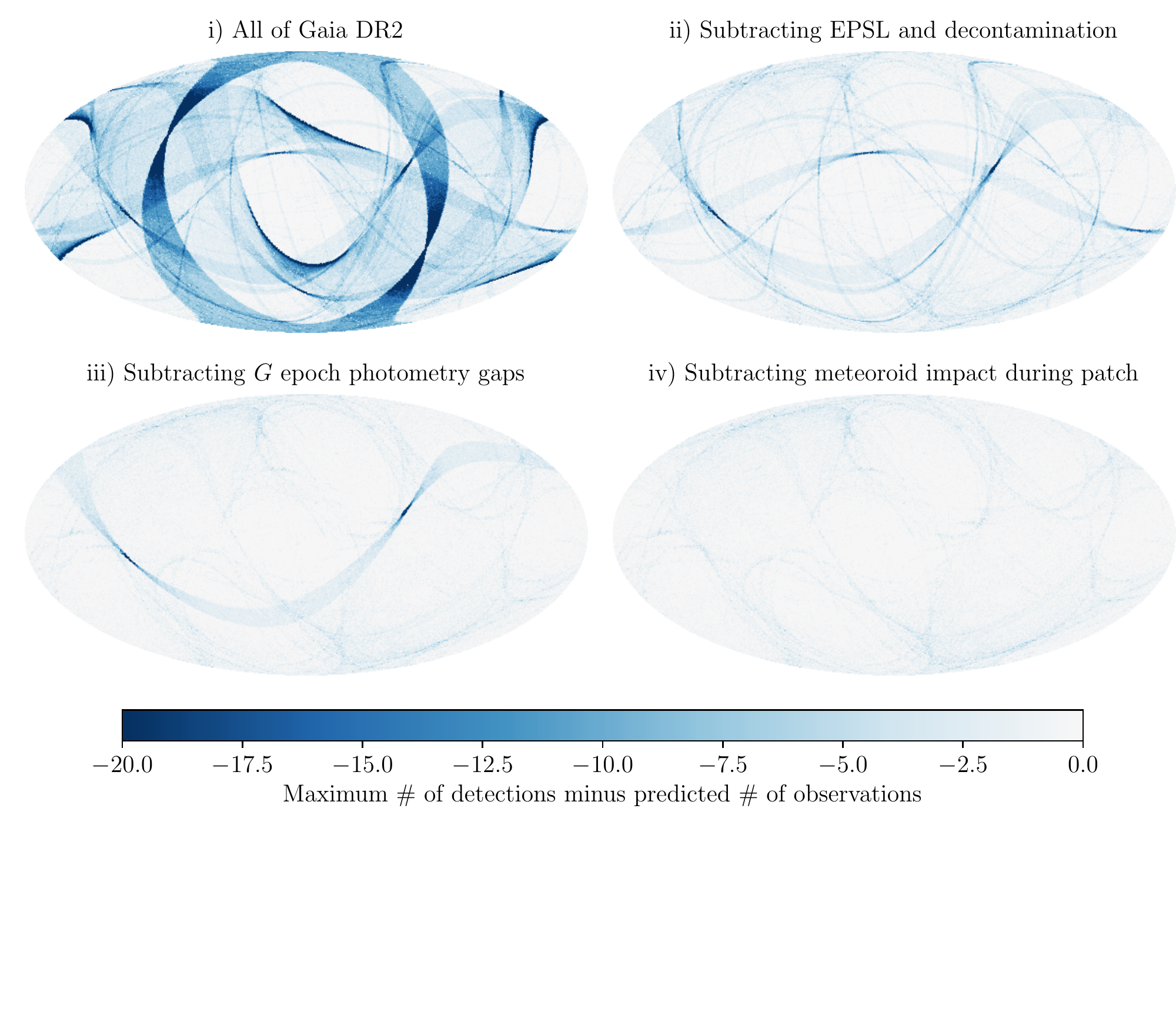}
	\caption{Maps in Galactic coordinates of the difference between the maximum number of astrometric detections \edits{of any source in each pixel of an $\textsc{nside}=128$ HEALPix map and the predicted number of observations of that pixel}, with each panel accounting for successively more gaps \edits{(the third panel includes the gaps shown in the second panel and the fourth panel includes the gaps shown in the third panel)}.}
	\label{fig:astrometry}
\end{figure*}

\begin{figure*}
	\centering
	\includegraphics[width=0.85\linewidth,trim=10 105 0 5, clip]{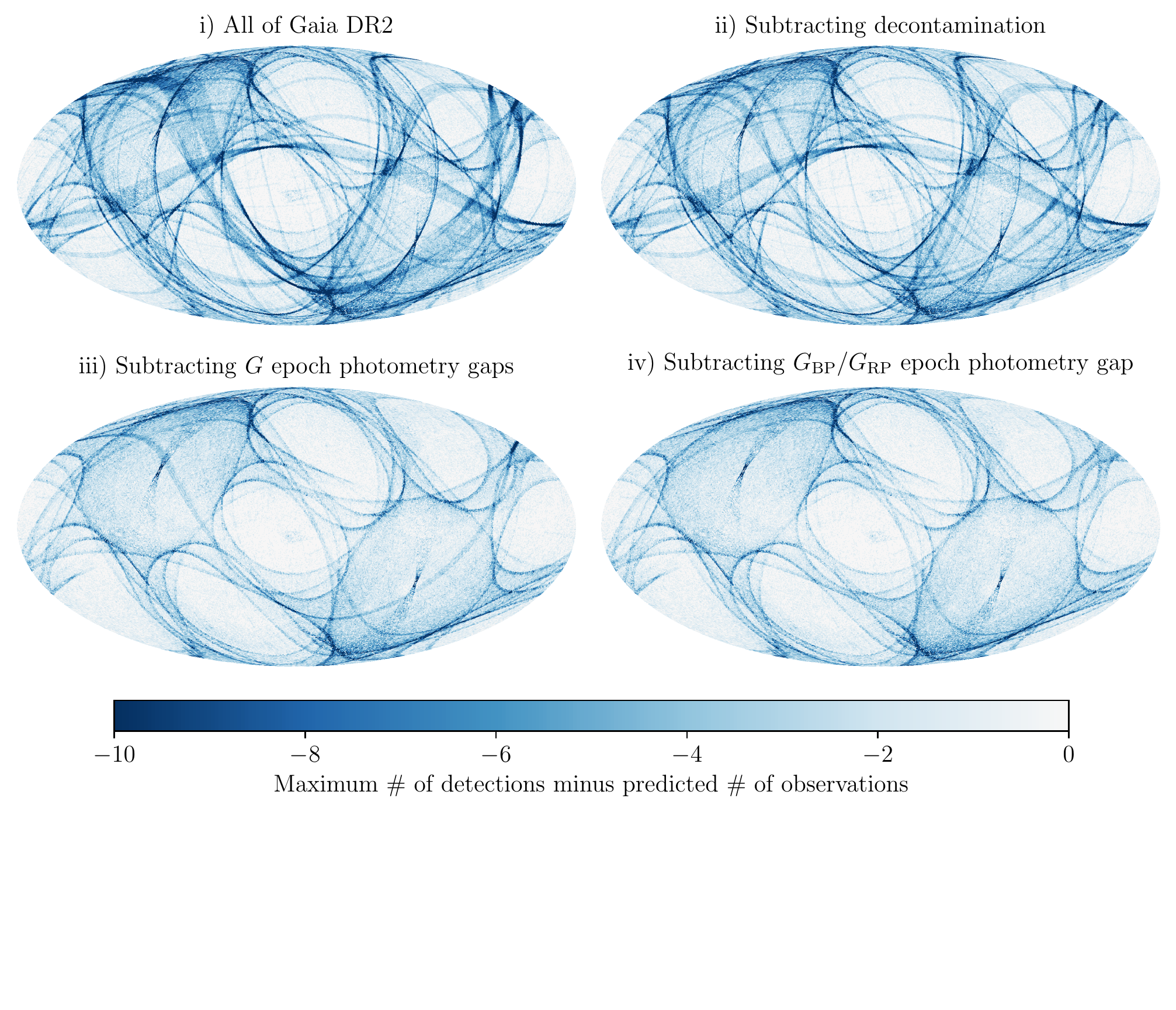}
	\caption{Same as Fig. \ref{fig:astrometry} but for the colour photometric detections.}
	\label{fig:photometry}
\end{figure*}

There are three relevant quantities when investigating gaps in the photometric pipeline: \textsc{phot}\_\textsc{g}\_\textsc{n}\_\textsc{obs} gives the number of individual astrometric CCD detections (of which there can be as many as nine during any one \edits{transit}), whilst \textsc{phot}\_\textsc{bp}\_\textsc{n}\_\textsc{obs} and \textsc{phot}\_\textsc{rp}\_\textsc{n}\_\textsc{obs} give the number of detections on the $G_{\mathrm{BP}}$ and $G_{\mathrm{RP}}$ CCDs respectively (of which there can only be one during each \edits{transit}). \citet{Riello2018} stated that the main gaps in the photometric pipeline were the decontaminations at $\mathrm{OBMT}=1316.492{-}1324.101\;\mathrm{rev}$ and $\mathrm{OBMT}=2330.616{-}2338.962\;\mathrm{rev}$. By inspecting the epoch photometry, we identified that any photometry taken during the EPSL prior to $\mathrm{OBMT}=1081\;\mathrm{rev}$ had also been discarded. These gaps together with the gaps identified from the epoch photometry in Sec. \ref{sec:epoch} completely define the gaps in \gaia data-taking with respect to the photometry. This is a tautological statement, because any gap in photometric data-taking would show up in the epoch photometry. To demonstrate the efficacy of this, we take $k$ to be the maximum of the \textsc{phot}\_\textsc{bp}\_\textsc{n}\_\textsc{obs} or \textsc{phot}\_\textsc{rp}\_\textsc{n}\_\textsc{obs} in each pixel and show maps of the difference between $k$ and $n$ in Fig. \ref{fig:photometry}. The residual scanning law pattern is stronger than in Fig. \ref{fig:astrometry} because of the apparently lower efficiency of the colour photometry detections as compared to the astrometric detections, meaning that the approximating the true number of observations by the maximum number of reported detections is a worse assumption for the colour photometry than for the astrometry.

\begin{figure}
	\centering
	\includegraphics[width=1.0\linewidth,trim=0 0 0 0, clip]{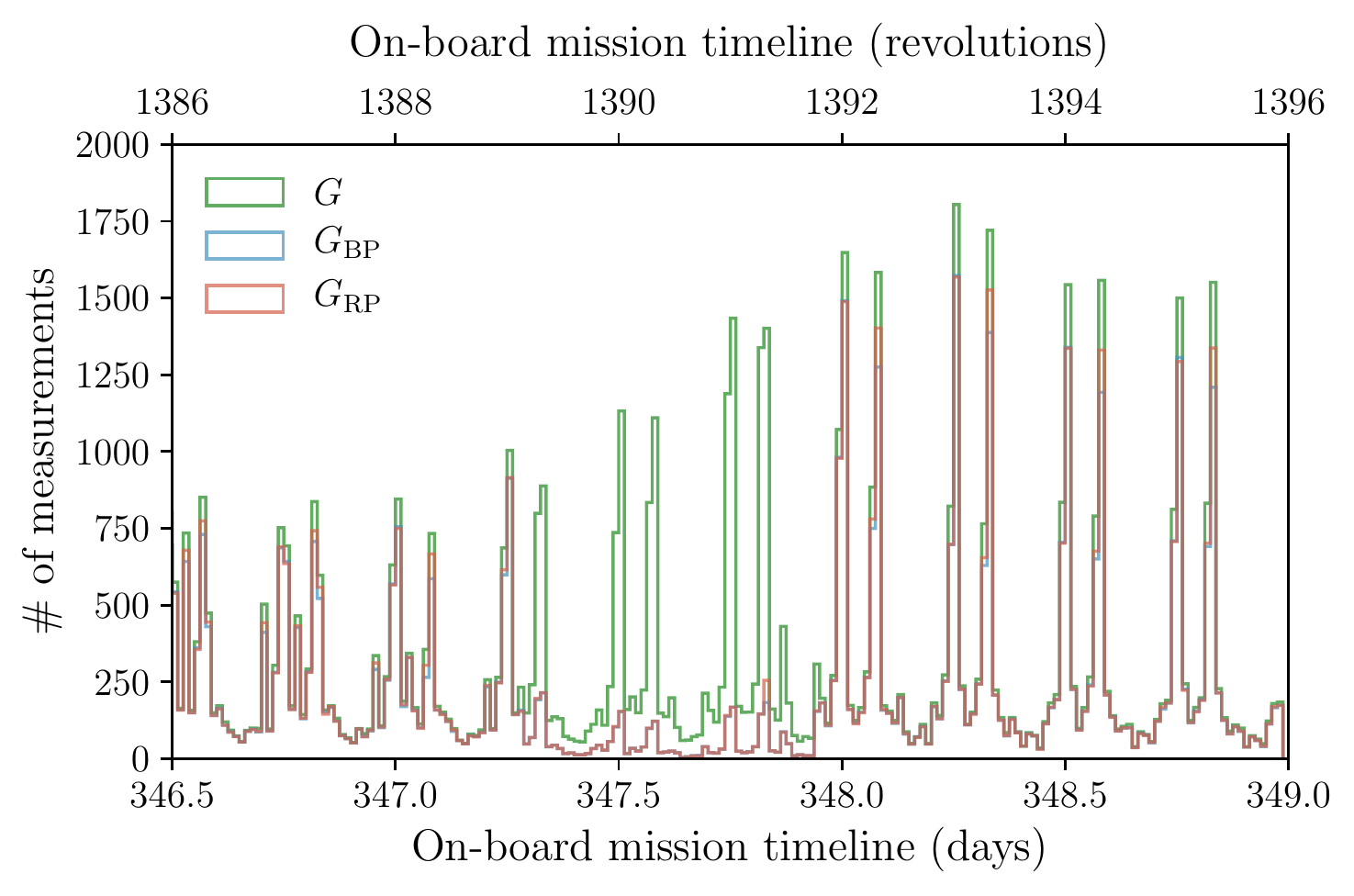}
	\caption{Histogram of the number of epoch photometric measurements of the 550,737 variable stars in each band over the period $\mathrm{OBMT}=1386{-}1396\;\mathrm{rev}$. There is a clear drop in the efficiency of the colour measurements \edits{beginning} at $\mathrm{OBMT}=1389.2\;\mathrm{rev}$, \edits{which is due to a   thunderstorm over Madrid preventing the down-link of data packets.}}
	\label{fig:slowdown}
\end{figure}

\begin{figure*}
	\centering
	\includegraphics[width=0.85\linewidth,trim=10 105 0 5, clip]{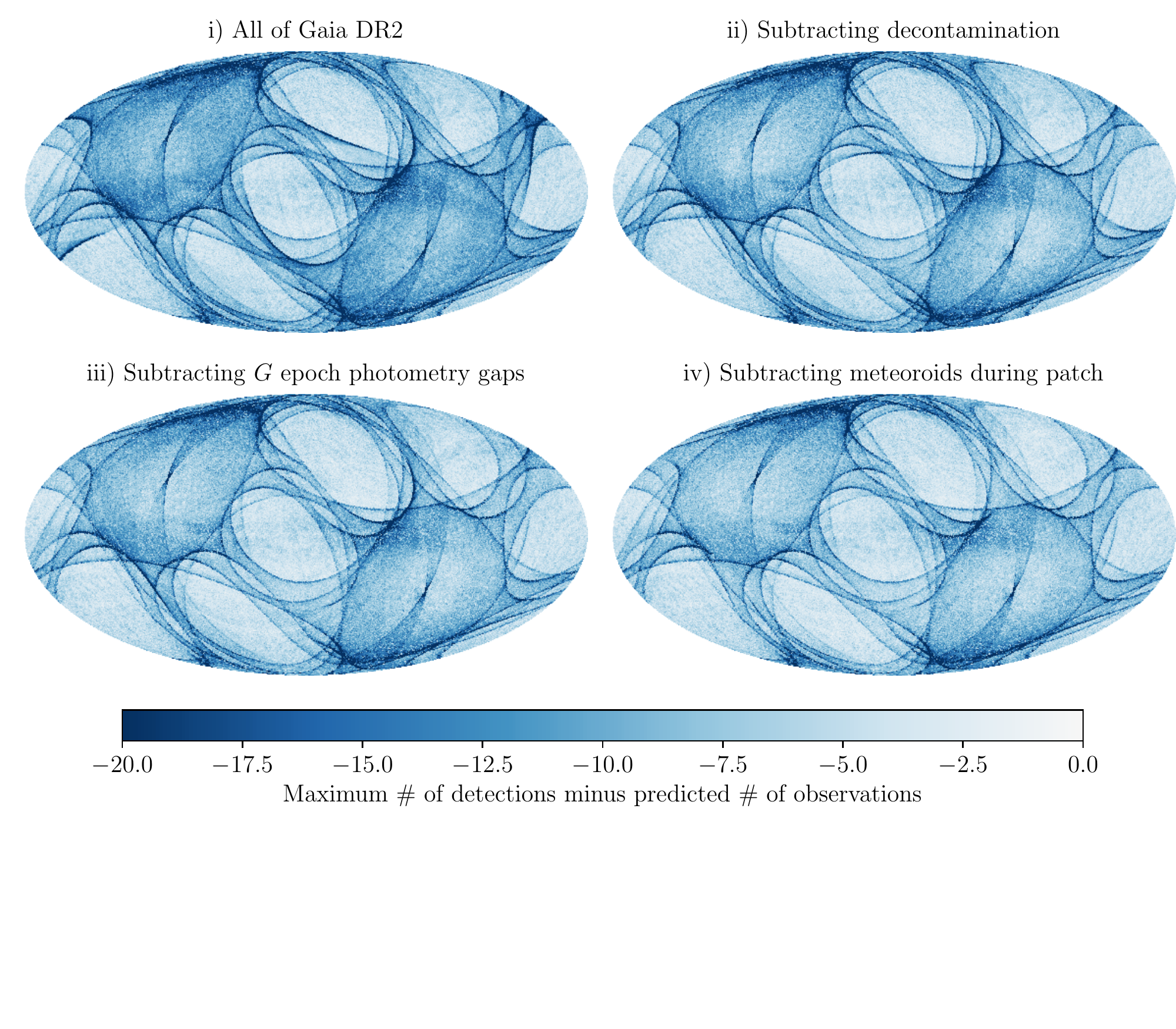}
	\caption{Same as Fig. \ref{fig:astrometry} but for the spectroscopic detections.}
	\label{fig:spectroscopy}
\end{figure*}

We note that there does appear to be another gap visible in the fourth panel of Fig. \ref{fig:photometry}, that crosses from the bottom right to the top middle. This gap is more visible in the fourth panel of Fig. \ref{fig:summary}. We were able to isolate this gap to the period $\mathrm{OBMT}=1389.2{-}1391.7\;\mathrm{rev}$. It is visible as a dip in the deviancy shown in Fig. \ref{fig:deviancy}. Curiously, this does not appear to be a genuine gap in photometric data-taking, but rather a drop in the efficiency of the colour photometry data-taking. We show in Fig. \ref{fig:slowdown} the rate of epoch photometric measurements in each band, and it is clear that the number of $G_{\mathrm{BP}}$ and $G_{\mathrm{RP}}$ measurements relative to $G$ measurements drops substantially over this period, but does not drop to zero. This explains why we did not identify this gap in Sec. \ref{sec:epoch}. We opted to not count this drop in efficiency as a gap \edits{for our purposes}, because some stars did receive colour photometric detections during this period. \edits{We queried the \gaia Helpdesk about this period and the following italicised text is an abridgement of their response.}

\edits{\emph{On Saturday evening 11 October 2014, around 18:55 UTC, while Gaia was transmitting data to the Cebreros ground station near Madrid, a thunderstorm developed over the Madrid sky-line and heavy rain started falling. As a result, contact with the spacecraft was lost until 19:04 UTC. However, during these 9 minutes, \gaia kept on transmitting its data to ground not knowing it would not be recorded. Whereas the bulk science data transmitted during this short interval was permanently lost, so-called critical auxiliary science data (ASD) packets that were lost were re-transmitted to ground the following day. This, however, was too late to use these packets in the regular, semi-live initial data treatment (IDT), which forms the first step in the astrometric and photometric (pre-)processing chains. As a result of the missing data, critical background information has been absent in the Gaia DR2 photometric processing for faint, one-dimensional windows ($G\geq 13$) for short stretches of time. The affected intervals (in OBMT revolutions) are 1389.7-1391.7 for row 2, 1389.2-1391.7 for row 3, 1389.2-1391.7 for row 4, 1389.2-1391.7 for row 5, 1389.2-1391.7 for row 6, and 1389.2-1391.3 for row 7; row 1 was not affected. The Gaia DR2 photometric calibration has "solved" the absence of background information by linearly interpolating between existing data (see Section 4.2 in \citealp{Riello2018}). This interpolation has, in this case, not worked perfectly and has failed to catch several straylight-induced peaks in the background. As a result, the photometry collected during these few revolutions is systematically biased and not reliable. In fact, the entire stretch from OBMT 1388.0 to 1392.0, which corresponds to the relevant calibration time interval, is indirectly affected by this issue. On the bright side: for Gaia (E)DR3, there is hope that this issue will be gone. Not only will gaps at IDT level have been fixed by the raw data reprocessing that has been undertaken, there has also been an update to the computation of the local background and this new feature should perform significantly better in periods with missing data.}}

\edits{We demonstrate the biased calibration of the photometry taken during this period in Appendix \ref{sec:dimstars}, where we show that almost all of the faintest stars in \gaia DR2 ($G>22$) can be attributed to spurious magnitude measurements taken during this time period.}

\subsubsection{Spectroscopy}

The number of individual \edits{detections of each source used in the DR2 spectroscopic pipeline} is given by the quantity \textsc{rv}\_\textsc{nb}\_\textsc{transits}, which in this section we will refer to as $k$. We show the difference between $k$ and $n$ in Fig. \ref{fig:spectroscopy}. We know from \citet{Sartoretti2018} that the two decontamination windows were removed, as in the other pipelines, but a precise interval was not given. There is no obvious way to obtain these periods quantitatively, unlike in the photometric case where we could turn to the epoch photometry. For this reason, we considered Fig. \ref{fig:deviancy} and identified three periods where the deviancy dropped substantially: two periods at $\mathrm{OBMT}=1316.492{-}1360.000\;\mathrm{rev}$ and $\mathrm{OBMT}=2324.9{-}2384.0\;\mathrm{rev}$ which appear to be the decontaminations, and an additional period at $\mathrm{OBMT}=2183.1{-}2226.8\;\mathrm{rev}$ which corresponds to the onboard software patching activity that upgraded the Video Processing Unit application software to version 2.8 between 24–28 April 2015\footnote{\url{https://gea.esac.esa.int/archive/documentation/GDR2/Introduction/chap_cu0int/cu0int_sec_release_framework/cu0int_sssec_spacecraft_status.html}}. We note that this software patch is related to the gap in the astrometry that occurred around the same time (see Fig. \ref{fig:deviancy}). In addition to these three gaps we also subtracted the gaps in the epoch photometry, which resulted in the relatively clear fourth panel of Fig. \ref{fig:spectroscopy}. The residual scanning law pattern is even stronger than in Figs. \ref{fig:astrometry} or \ref{fig:photometry}, because of the lower \edits{effective} efficiency of the radial velocity detections. We note that the Galactic plane is also visible in this plot as crowding is more effective at preventing spectroscopic detections than astrometric or photometric ones. The only clear remaining feature in the fourth panel is a ring that occurred during the EPSL and is visible in Fig. \ref{fig:deviancy} as a dip shortly after the start of nominal observations. We opted not to account for this period, both because this is a quite short gap and because we cannot know whether this is a genuine gap or simply a drop in efficiency. We note that the lower efficiency of the \gaia spectroscopic detections means that it is much more difficult to identify shorter gaps than in the astrometric or photometric cases.

\subsection{Summary}

\begin{figure*}
	\centering
	\includegraphics[width=1.0\linewidth,trim=10 95 0 5, clip]{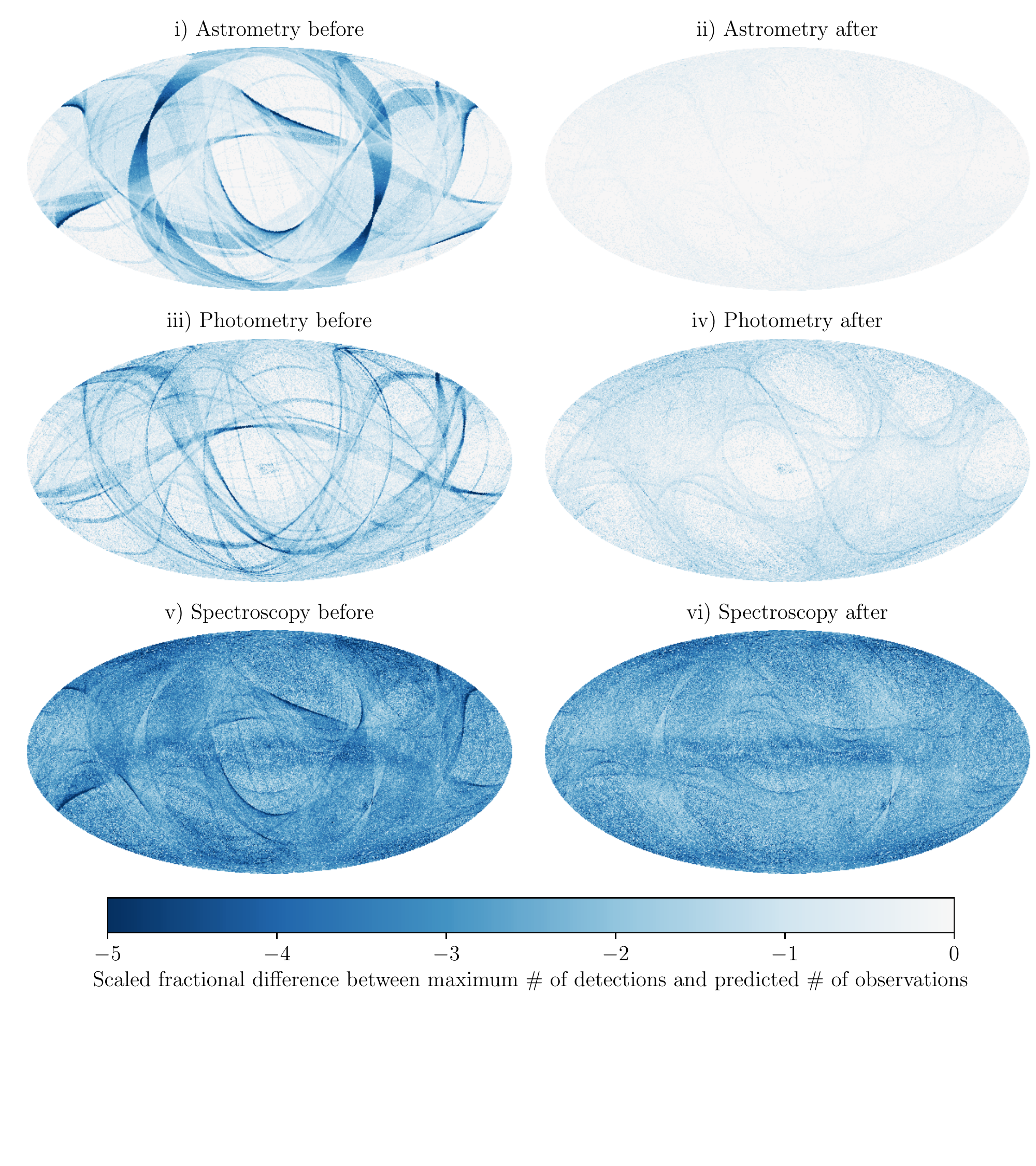}
	\caption{The fractional difference between the maximum number of detections and the predicted number of observations, before and after removing the gaps in \gaia data-taking identified in this work. Both plots in each row are scaled by the interquartile range of the plot in the left-hand column.}
	\label{fig:summary}
\end{figure*}

As an overview of the gaps we have identified, we show in Fig. \ref{fig:summary} before and after panels of the relative fractional error of the number of observed detections versus predicted number of observations in each pipeline, scaled to the interquartile range of the before panel. Much of the structure visible in the left-hand column has been successfully removed. We conclude that we have identified the overwhelming majority of the \edits{periods in time where that did not contribute any detections to the \gaia DR2 data products}.

\section{Spurious and duplicated observations}
\label{sec:duplicates}

\begin{figure}
	\centering
	\includegraphics[width=1.\linewidth,trim=0 0 0 0, clip]{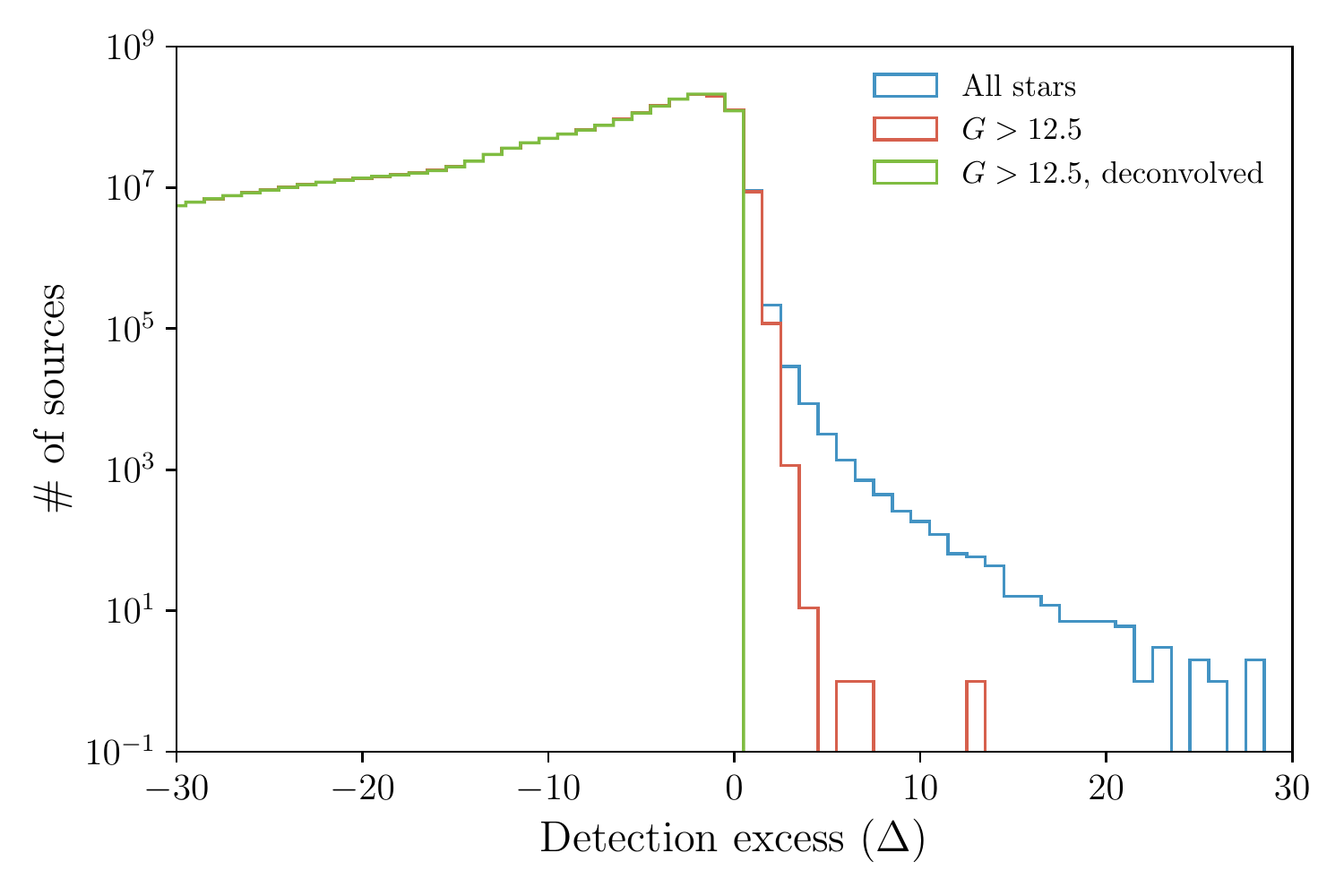}
	a) Distribution of differences between detection and predicted observations.\vspace{0.5cm}
	
	\includegraphics[width=1.\linewidth,trim=0 0 0 0, clip]{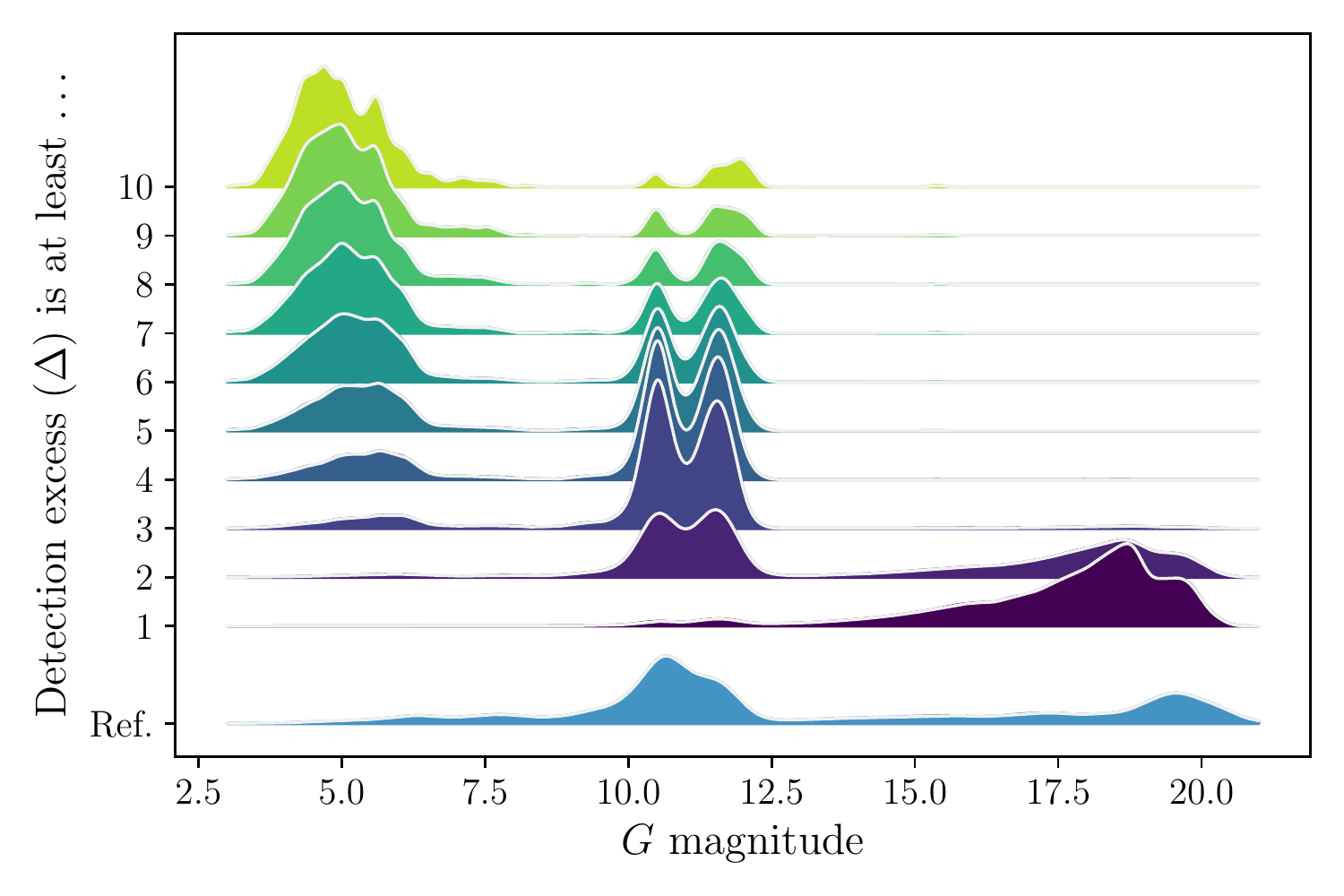}
	b) Magnitude distributions of stars with excess detections.\vspace{0.5cm}
	
	\caption{There are sources with more reported detections used by the astrometric pipeline than we predict could have been taken by \gaia during DR2, e.g. there is an detection excess ($\Delta$). \textbf{Top:} We show the number distribution of stars by their detection excess for all stars in \gaia DR2 (blue) and for the subset with $G>12.5$ (red), demonstrating that the long tail to large $\Delta$ is almost exclusively made up of stars brighter than $G<12.5$. There is a roughly $0.5$ standard deviation error in our prediction of the number of observations of a source, and we demonstrate by applying Richardson-Lucy deconvolution with the appropriate kernel that the faint ($G>12.5$) stars with detection excesses are entirely due to us under-predicting the number of observations. \textbf{Bottom:} Magnitude distributions of the sources with $\Delta \geq n$ for increasing $n$ showing that bright stars dominate for $\Delta\geq3$. The reference distribution shows the spurious detections we identified in the epoch photometry, demonstrating that the cause of the excess detections are spurious duplicate detections which are being wrongly counted as detections of genuine sources.}
	\label{fig:duplicates}
\end{figure}

\edits{A key assumption of the previous section is that each \gaia observation of a source results in only one detection, however there are multiple ways that this assumption can have broken (\gaia Helpdesk, private communication):
\begin{itemize}
    \item The diffraction of light from bright stars (or even planets in the Solar system) can cause spurious secondary images out to several arcminutes. If these are nearby on the sky to a genuine source then the secondary image of the bright star can be wrongly matched to that source.
    \item Marginally resolved binary stars with separations of less than $1\;\mathrm{arcsec}$ can be detected as one or two sources depending on the scanning angle of each observation. These sources can also produce spurious detections where the along-scan diffraction wing of one component crosses the across-scan diffraction wing of the other.
    \item Some genuinely single sources can generate a double detection on the \gaia CCDs. This issue is known to vary between CCDs and to depend on the instantaneous contamination, focus state and stellar density in the field of view.
\end{itemize}
Identifying and removing these spurious detections is a challenging task, particularly because these scenarios can occur in combination with each other. Almost all of these spurious detections have been filtered out by the \gaia pipeline, however in this section we identify remaining contamination that has affected both the astrometric and variable star classification pipelines.}

\edits{The smoking gun that alerted us to the issue of duplicates was the source Gaia DR2 1633302743286498176, a bright $G=8.77$ star at the North Ecliptic Pole. This star has an impossibly large 361 matched observations, i.e. when all of the individual detections made by \gaia during the 22 months of DR2 were being merged into sources, 361 unique detections were merged into this one source. As shown in Fig. \ref{fig:map}, no source should have more than 264 observations, implying that almost one hundred of the detections of this star are spurious duplicates. We note that the column \textsc{matched\_observations} in the \gaia DR2 source catalogue has not been filtered for duplicates, with each of the astrometric, photometric, spectroscopic and variable star pipelines being left to make their own cuts.}

\edits{To gauge the efficacy of the cuts applied by the astrometric pipeline, we computed the difference between the predicted number of astrometric observations (accounting for the gaps detailed above) of each source in \gaia DR2 and the reported number of astrometric detections in the column \textsc{astrometric\_matched\_observations}. We show a histogram of this `detection excess' $\Delta$ in the top panel of Fig. \ref{fig:duplicates}. We stress that not all observations of a source result in detections, and thus that \textsc{astrometric\_matched\_observations} is a lower bound on the true number of observations. While almost all the sources have no excess of detections, there is a tail of sources with excesses of as much as thirty. Excesses that large cannot be explained by the error in our prediction of the number of observations and so these sources must be suffering from duplicate detections. In the bottom panel of Fig. \ref{fig:duplicates} we show empirical magnitude distributions (smoothed with a $0.05\;\mathrm{mag}$ Gaussian kernel) of all sources with more excess detections than a series of increasing thresholds. At small detection excesses the distribution is dominated by faint sources, whilst sources with higher detection excess are predominantly bright. We conclude that most of the sources with small detection excesses are simply sources where almost every observation resulted in a detection but we have under-predicted the number of observations. These sources are approximately a random draw from \gaia DR2 and so their magnitude distribution approximately matches that of \gaia DR2 as a whole. The stars with large detection excesses, however, have a peculiar magnitude distribution with peaks at 5, 10 and 12 which resembles the magnitude distribution\footnote{We identified spurious duplicate detections in the epoch photometry by looking for two observations of a source spaced less than $1\;\mathrm{hour}$ apart. We assumed one of these detections was a genuine observation of the source whilst the other was spurious, and assumed that whichever flux measurement was further from the median flux of that source was the spurious one.} of the spurious duplicate detections we identified in the epoch photometry (shown for reference in Fig. \ref{fig:duplicates}).}

\edits{Motivated by the predominantly bright nature of the stars with many excess detections, we show in the top panel of Fig. \ref{fig:duplicates} the histogram of detection excesses for stars fainter than $G>12.5$. This cut eliminates almost all of the large excess stars, apart from \gaia DR2 5097875186758444800 $(G=15.41,\Delta=7)$, \gaia DR2 5866952440466180608 $(G=15.38,\Delta=13)$ and \gaia DR2 6381503193406409088 $(G=15.12,\Delta=6)$. There are duplicate observations at these magnitudes in the epoch photometry, although they are rarer than the duplicate observations at brighter ($G<12.5$) and fainter ($G>18$) magnitudes. We note that we will not uncover faint sources with duplicate observations in our \textsc{astrometric\_matched\_observations} test, because faint sources have a low efficiency in turning observations into detections and so even a significant number of spurious detections will not be sufficient to cause them to have a positive detection excess. These three sources are overwhelmingly likely to be suffering from a large number of duplicate observations based on their large astrometric reduced unit weight error values ($\mathrm{RUWE}>10$) and so we discard these three sources for the next part of our analysis.}

\edits{If we are correct that the excess detection distribution of the faint stars can be attributed to errors in our prediction of the number of observations, then the histogram of stars with $G>12.5$ in the top panel of Fig. \ref{fig:duplicates} can be considered to be the result of convolving the true excess detection distribution (which is zero for $\Delta>0$) with the Skellam error distribution we identified in Sec. \ref{sec:predicting}. To demonstrate this  we applied the Richardson-Lucy deconvolution algorithm to the histogram under the assumption that the underlying true distribution had been convolved with our Skellam error distribution, and show the result as the green line in the top panel of Fig. \ref{fig:duplicates}. Our estimate of the true underlying distribution is consistent with no stars having an excess of detections over the number of predicted observations, and thus that the excess detection distribution of the faint sources is due entirely to small errors in our prediction of the number of observations.}

\edits{Returning to the 492,415 bright ($G<12.5$) stars with excess detections ($\Delta>0$), we investigated whether the spurious detections had impacted their \gaia DR2 data products. We selected the 6,517 stars with $G<12.5$ and $\Delta \geq5$ as our sample and randomly drew an equal-sized control group from the set of stars with $G<12.5$ and $\Delta \leq0$, with the random draw weighted such that there were equal numbers of stars in log-spaced $G$ band flux bins. Relative to the control, our sample of stars with $\Delta\geq5$ were 174\% more likely to have $\varpi/\sigma_{\varpi}<0$ and 89\% more likely to have $\mathrm{RUWE}>1.4$, suggesting some correlation between a star having spurious detections and it being an astrometric outlier in the DR2 data products. Spurious detections can result in increased astrometric noise because they will be in the wrong location on the sky compared to true detections of the source. However, we note that the astrometric pipeline explicitly down-weights detections which it finds to not be in agreement with the majority, with the result that only 1\% of the stars with $\Delta\geq5$ have $\varpi/\sigma_{\varpi}<0$ and 15\% have $\mathrm{RUWE}>1.4$. Additionally, our sample stars were 57\% more likely to be classified as variable. Because all duplicates were removed from the variable star analyses \citep[see Sec.~7.2.3 of][]{Eyer2018}, which should remove any bias these duplicates would have caused in the photometry, this finding suggests that intrinsically variable stars are more affected by the duplicate observation issue. This is corroborated by \citet{Mowlavi2018} who found that the DR2 Long Period Variables with very large variability amplitudes and extreme colours had an overabundance of spurious detections.}

\edits{We note that, while duplicate observations appear to not have affected the astrometry of these bright sources, they may have a greater impact on the astrometry of fainter sources, where they could make up a greater proportion of the detections of the source. We further note that the most important cut which decides whether a source is in \gaia DR2 is $\textsc{astrometric\_matched\_observations}\geq5$ \citep{Lindegren2018} and we have shown in this section that that quantity contains at least some duplicate detections, implying that the phenomenon of duplicate detections has to some degree distorted the selection function of the \gaia DR2 source catalogue.}

\section{Conclusions}
\label{sec:conclusion}

\edits{The \gaia mission has broken astrometric, photometric and spectroscopic records with its second data release, but if we want to interpret what \gaia has seen then we need to fully appreciate the spinning-and-precessing way that \gaia looks at the sky. While the nominal \gaia scanning law is known and public, it can differ by up to $30\;\mathrm{arcsec}$ from the true attitude of \gaia. Furthermore, there are undisclosed times when the astrometric, photometric or spectroscopic data acquired by \gaia was not used in the published data products. In this work we have found these gaps by identifying breaks in the epoch photometry and by looking for consecutively scanned locations on the sky where the predicted number of observations was greater than the observed maximum number of detections. These gaps will be important to bear in mind whenever the number of detections is used as a quality cut, which we note is \emph{essentially every paper using \gaia DR2 data}. The gaps will be especially important in applications that depend on the individual \gaia epoch measurements; for instance, seven of the thirty astrometric microlensing events predicted by \citet{Kluter2018} fall in gaps we have identified for the astrometric pipeline. The astrometric measurements of these microlensing events may be recoverable with the improved pipelines that will be used for future \gaia data releases, but certainly some will not (e.g. if the gap is due to telemetry loss). On the way to that result, we inferred the most accurate publicly available determination of the \gaia scanning law, again using the DR2 variable star epoch photometry. In the future papers of this series we will use the predicted number of observations -- corrected for gaps in the data-taking and using our precision determination of the scanning law -- to infer unbiased selection functions for the \gaia-verse of astrometric, photometric and spectroscopic catalogues. By inferring both the gaps in \gaia's data-taking and where \gaia was looking with time, we have made it possible to truly exploit the times when \gaia's eye was on the sky.}

\section*{Postface}
\edits{At various points in the development of this work we queried the \gaia Helpdesk about possible gaps which did not contribute to \gaia DR2. Our questions inspired an effort within DPAC to publish an official list of those gaps (\url{https://www.cosmos.esa.int/web/gaia/dr2-data-gaps}), which was made public on the day that this paper was resubmitted after our first revisions. The official lists of astrometric and spectroscopic are more accurate than ours and so should always be used in preference to ours. There is no definitive list of the short photometric gaps and it is thus possible that some of the gaps we identified in the epoch photometry are not in the official lists.We stress that the methods we used to identify the gaps in Sec. \ref{sec:methodology} will be applicable to future data releases to identify `outlying' time intervals or gaps, especially if the official gaps will be published after the data releases themselves.}

\section*{Acknowledgements}
\edits{The authors are grateful to the \gaia Data Processing and Analysis Consortium for their earnest efforts to answer our tricky questions through the go-between of the \gaia Helpdesk. We particularly thank Jos de Bruijne and Timo Prusti for fast and detailed responses to queries on the spacecraft's operations, and the team behind the \gaia Observation and Forecasting Tool for generating the publicly available nominal scanning law. We would like to further thank the anonymous reviewer for inspiring us to be more ambitious than in our original manuscript.} DB thanks Magdalen College for his fellowship and the Rudolf Peierls Centre for Theoretical Physics for providing office space and travel funds. AE thanks the Science and Technology Facilities Council of the United Kingdom for financial support. This work has made use of data from the European Space Agency (ESA) mission \gaia (\url{https://www.cosmos.esa.int/gaia}), processed by the \gaia
Data Processing and Analysis Consortium (DPAC,
\url{https://www.cosmos.esa.int/web/gaia/dpac/consortium}). Funding for the DPAC
has been provided by national institutions, in particular the institutions
participating in the \gaia Multilateral Agreement.




\bibliographystyle{mnras}
\bibliography{references} 



\FloatBarrier

\appendix

\section{Inferring the roll-and-pitch of Gaia}
\label{sec:pitchandroll}

\begin{figure}
	\centering
	\includegraphics[width=1.0\linewidth,trim=0 0 0 0, clip]{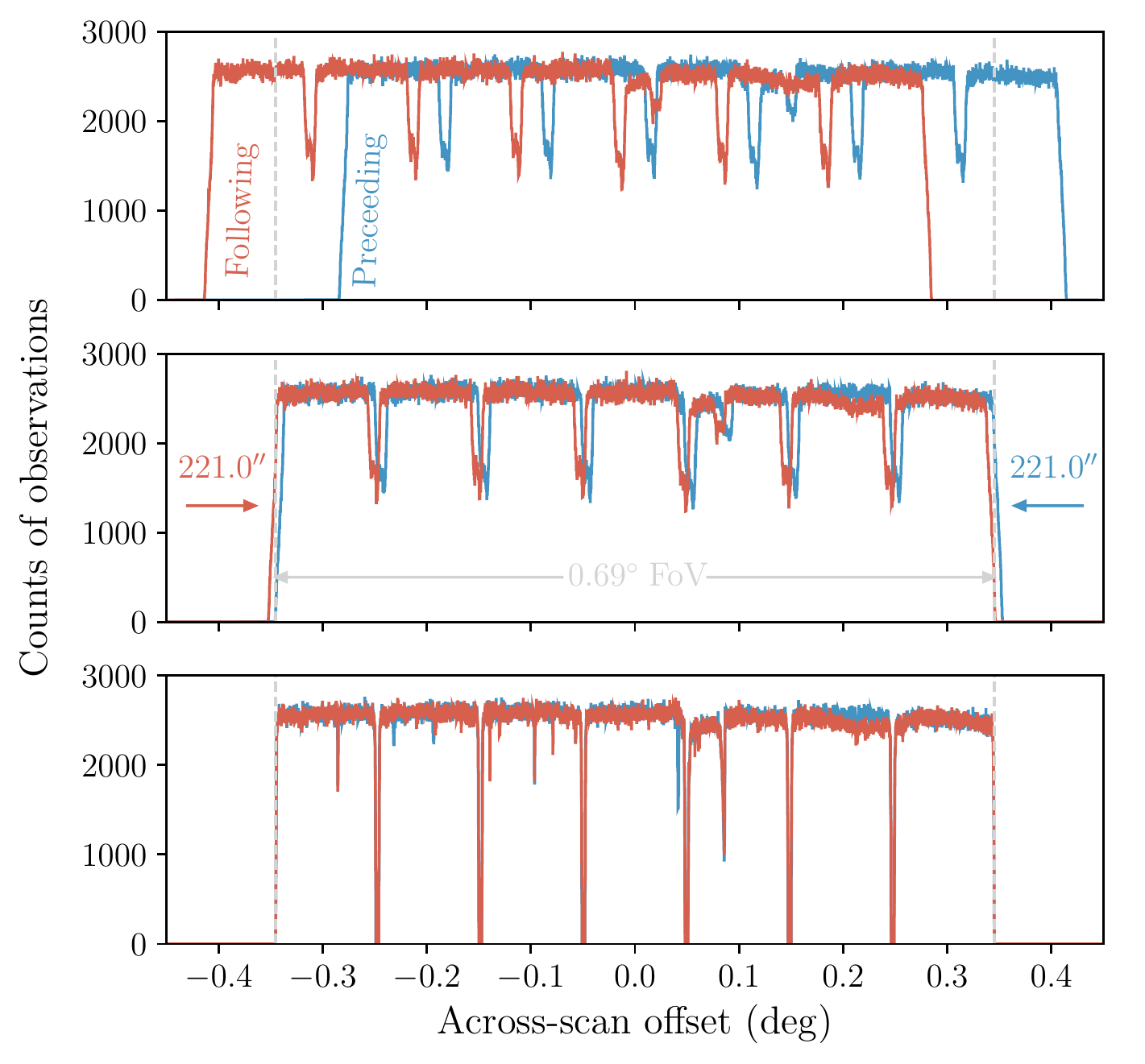}
	\caption{Across-scan distances of all the sources with epoch photometry from the centre of the focal plane at their times of observation. \textbf{Top:} Across-scan offsets without correcting for the offset origins of the two fields of view and using the DPAC-published nominal scanning law. The grey lines indicate the $0.69^{\circ}$ wide \gaia focal plane. \textbf{Middle:} Across-scan offsets corrected using the nominal offsets given by \citet{Lindegren2012}. \textbf{Bottom:} Across-scan offsets additionally corrected using our improved determination of the \gaia scanning law.}
	\label{fig:deltahists}
\end{figure}

\begin{figure*}
	\centering
	\includegraphics[width=1.0\linewidth,trim=0 0 0 0, clip]{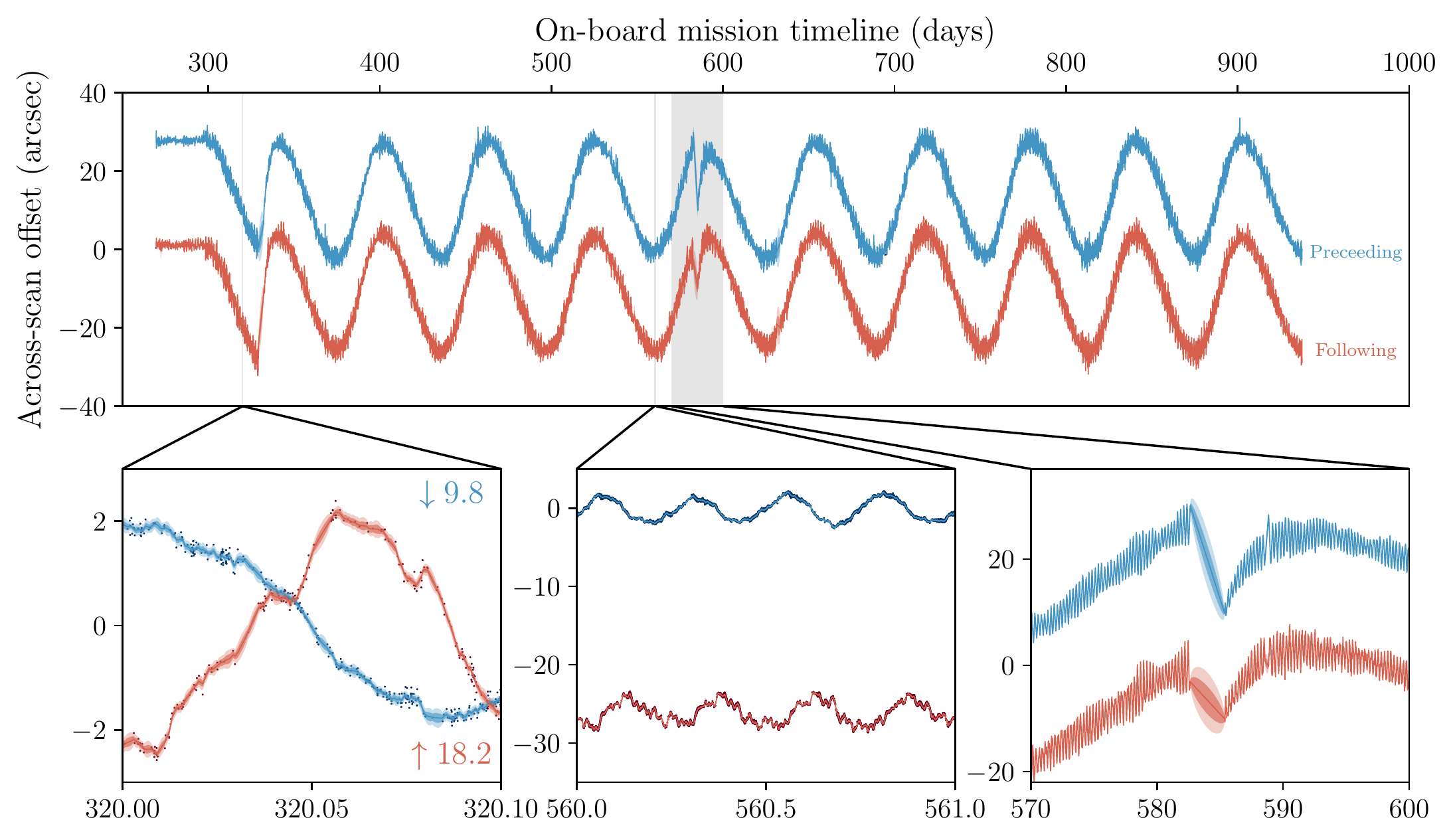}
	\caption{Our posterior model for the across-scan offsets of the two fields of view with the associated one and two sigma regions. This model was fit to the across-scan offsets of all the sources with epoch photometry at their times of observation and these data points are shown in the bottom left and bottom middle panels. The bottom row of panels zoom in to the indicated time periods in the top panel, with the curves in the bottom left panel having been linearly offset by the indicated values to maximise the dynamic range of the y-axis.}
	\label{fig:attitude}
\end{figure*}

This appendix is concerned with the geometry of mapping locations of stars in the field-of-view on the sky to their location on the focal plane. \gaia rotates every 6 hours in the plane of the great circle defined by the two FoV centres, which are separated by $\Gamma_c = 106.5^{\circ}$, with that great circle precessing with a 63 day period to give full sky coverage. The instantaneous across-scan position of a source at any point in time is simply the angular distance of the source from the appropriate great circle. The \gaia DPAC has published a nominal scanning law which gives the commanded locations of the FoV centres at 10 second intervals over the course of DR2, thus allowing us to calculate the nominal across-scan positions of sources with time. For each $G$-band epoch photometric observation discussed in the main text, we estimated the across-scan position $B$ of the source at the time it was observed by a linear weighted average of the across-scan position at the nearest two scanning law time steps. We show a histogram of these across-scan distances for the preceeding and following FoVs in the top panel of Fig. \ref{fig:deltahists}. The seven CCD rows are visible as tophat-like shapes with gaps between them. That the edges of these tophats are not perfectly straight is due to small difference between the commanded nominal scanning law that we are using and the true scanning law carried out by \gaia, which can deviate\footnote{\url{https://www.cosmos.esa.int/web/gaia/scanning-law-pointings}} from each other by up to $30\;\mathrm{arcsec}$. The tops of the tophats are not flat due to pixel-to-pixel variations in each CCD. The technical design of \gaia causes each FoV to have an across-scan offset such that the FoV centres do not lie along the middle of the focal plane, which is visible in the Top Panel of Fig. \ref{fig:deltahists} as an offset between the two FoVs. The across-scan origin of each FoV is a critical number to know when predicting when sources were observed by \gaia, because it moves the predicted projection of the FoVs by hundreds of arcseconds. \citet{Lindegren2012} gives the authoritative account of the geometry of \gaia and states that the across-scan origins are at $\pm221\;\mathrm{arcsec}$, and we show the result of applying these offsets in the Middle Panel of Fig. \ref{fig:deltahists}. The remaining residual offset can be explained by the difference between the nominal and true scanning laws mentioned above. This attitude difference also causes sharp features in the across-scan distribution to be blurred out, with the gaps between the CCDs appearing ten times wider than they are in reality. Errors in the predicted across-scan locations of sources on the focal plane of as much as $30\;\mathrm{arcsec}$ will hinder our ability to precisely determine which sources are observed at each point in time, which will have implications for the selection functions we derive in future papers. We therefore opted to infer a correction to the published nominal scanning law using the \gaia DR2 epoch photometry.

We defined a Cartesian coordinate frame at every instant in time such that the fields of view lie at $(x,y,z)=(\cos{\gamma_c},\pm\sin{\gamma_c},0)$, where we have defined the shorthand $\gamma_c=\Gamma_c/2$. We consider extrinsic rotations about the $x$ (roll) and $y$ (pitch) axes,
\begin{align}
    R(\theta_x,\theta_y)&=R_x(\theta_x)R_y(\theta_y) \\ 
    &= \begin{bmatrix}
1 & 0 & 0 \\
0 & \cos \theta_x &  -\sin \theta_x \\[3pt]
0 & \sin \theta_x  &  \cos \theta_x \\[3pt]
\end{bmatrix} \begin{bmatrix}
\cos \theta_y & 0 & \sin \theta_y \\[3pt]
0 & 1 & 0 \\[3pt]
-\sin \theta_y & 0 & \cos \theta_y\\
\end{bmatrix} \\
&= \begin{bmatrix}
\cos \theta_y & 0 & \sin \theta_y \\
\sin \theta_x \sin \theta_y & \cos \theta_x &  -\sin \theta_x \cos \theta_y \\[3pt]
-\cos\theta_x \sin\theta_y & \sin \theta_x  &  \cos \theta_x \cos \theta_y \\[3pt]
\end{bmatrix} \label{eq:rotmatrix}\\
&\approx \begin{bmatrix}
1 & 0 & \theta_y \\
0 & 1 &  -\theta_x \\[3pt]
-\theta_y & \theta_x  &  1 \\[3pt]
\end{bmatrix},
\end{align}
where in the final line we used the small angle approximations $\cos\theta\approx1$ and $\sin\theta\approx\theta$ and dropped second order terms. The two fields of view are rotated to 
\begin{equation}
    R(\theta_x,\theta_y)\begin{bmatrix}\cos{\gamma_c} \\ \pm\sin{\gamma_c} \\ 0 \end{bmatrix} = \begin{bmatrix}\cos{\gamma_c} \\ \pm\sin{\gamma_c} \\ -\theta_y\cos{\gamma_c}\pm \theta_x\sin{\gamma_c}\end{bmatrix}.\label{eq:rotfov}
\end{equation}
The non-zero $z$-component corresponds to a shift in the across-scan direction of the original frame. Noting the signs of these shifts, we see that rotations about the $y$-axis cause both FoVs to be offset in the same direction while rotations about the $x$-axis cause the two FoVs to be offset in opposite directions, and thus that both rotations are necessary to explain the across-scan shifts of stars in the two FoVs.

Let $\hat{B}$ be the predicted across-scan location of the star in the original frame, then the corrected across-scan location of the star in the rotated frame is 
\begin{equation}
    B(\theta_x,\theta_y|\hat{B}) = \hat{B} - \left(-\theta_y\cos{\gamma_c}\pm \theta_x\sin{\gamma_c}\right),
\end{equation}
depending on which field of view the star is observed in and where the roll and pitch are applied to the frame defined by the nominal scanning law at the time $t$ of the observation of the source. $B(\theta_x,\theta_y|\hat{B})$ is therefore our model of the across-scan location of the source. The observational data we will be comparing to is hidden in the $\textsc{transit\_id}$ column of the epoch photometry table, from which we can extract the field of view $F$, CCD $C$ and pixel $P$ of the observation when it was first acquired on the AF1 CCD (i.e. the first column of the astrometric field of CCDs). After correcting these indices such that they are zero-indexed, we can estimate the across-scan location of the observation as
\begin{equation}
    \tilde{B}(F,C,P) = \Delta_f(1-2F)+\Delta_c(C-3)+\Delta_p(P-1966/2), \label{eq:model}
\end{equation}
where $\Delta_f=220.9979\;\mathrm{arcsec}$ is the magnitude of the across-scan offset of each of the fields of view, $\Delta_c=356.5435\;\mathrm{arcsec}$ is the across-scan size of each CCD, and $\Delta_p=0.1768\;\mathrm{arcsec}$ is the across-scan size of each pixel. These were calculated assuming that the offset between the fields of view is $\pm37.5\;\mathrm{mm}$, that the pixels are $30\;\mu\mathrm{m}$ across and that the spacing between CCDs in the across-scan direction is $1.52\;\mathrm{mm}$, and these lengths were converted to angles assuming a focal length of $35\;\mathrm{m}$. The precision of this estimate of the across-scan location of each observation is limited by the size of the pixels, and thus the true across-scan location of the observation will have been somewhere in the range $\tilde{B}\pm\Delta_p/2$. The likelihood of the observation given our model is therefore $\tilde{B}\sim \operatorname{Uniform}\left(B(\theta_x,\theta_y)-\Delta_p/2,B(\theta_x,\theta_y)+\Delta_p/2\right)$. Each of the numbers $(\Delta_f,\Delta_c,\Delta_p)$ has some level of uncertainty due simply to the finite precision of the numbers provided in the DPAC papers, and there may be further uncertainty due to shifts and contractions of \gaia on its journey from the clean room to L2. We include two additive correction factors $(\delta_c,\delta_p)$ to our model which allow us to account for these uncertainties, noting that we do not have a correction factor $\delta_f$ because changes in the offset between the fields-of-view are entirely degenerate with rolls of \gaia.

\begin{table*}
    \centering
    \caption{Schema of the electronic file (\url{https://doi.org/10.7910/DVN/OFRA78}) in which we have published our determination of the \gaia scanning law during DR2, which we note is identical in format to the \gaia DPAC nominal scanning law file (\url{https://www.cosmos.esa.int/web/gaia/scanning-law-pointings}).}
    \label{tab:scanning law}
    \begin{tabular}{llr}
\hline Column                                       & Description                                                                         & Example entry         \\
\hline\textsc{JulianDayNumberRefEpoch2010TCB@Gaia}          & Time at \gaia                                  & 1666.4384953703 \\
\textsc{JulianDayNumberRefEpoch2010TCB@Barycentre\_1} & Time at barycentre for sources in the preceding FoV & 1666.4351984748 \\
\textsc{JulianDayNumberRefEpoch2010TCB@Barycentre\_2} & Time at barycentre for sources in the following FoV & 1666.4419269210 \\
\textsc{ra\_FOV\_1(deg)}                              & Right ascension in degrees of the preceding FoV                           & 152.1064885273  \\
\textsc{dec\_FOV\_1(deg)}                             & Declination in degrees of the preceding FoV                               & -29.9145450844  \\
\textsc{scanPositionAngle\_FOV\_1(deg)}               & Scan position angle in degrees of the preceding FoV                       & 26.5781113809   \\
\textsc{ra\_FOV\_2(deg)}                              & Right ascension in degrees of the following FoV                           & 4.5898138374    \\
\textsc{dec\_FOV\_2(deg)}                             & Declination in degrees of the following FoV                               & -36.9854274849  \\
\textsc{scanPositionAngle\_FOV\_2(deg)}               & Scan position angle in degrees of the following FoV                       & 150.9550696213 \\ \hline
\end{tabular}
    
\end{table*}

At each epoch observation $i$ we have the times of the observation $t_i$, the FoV $F_i$, CCD $C_i$ and pixel $P_i$, and a flag $s_i$ which takes values of $\pm1$ for observations in the preceeding ($-1$) and following ($+1$) FoVs. For fixed values of the parameters $(\delta_c,\delta_p)$, we can estimate the across-scan location of each observation $\tilde{B}_i$ using Eq. \ref{eq:model}, noting that we take $(\Delta_f,\Delta_c,\Delta_p)\rightarrow(\Delta_f,\Delta_c+\delta_c,\Delta_p+\delta_p)$. We also have the predicted across-scan location given the nominal scanning law $\hat{B}_i$. We choose to model the time-variation of the rolls $\theta_x(t_i)$ and pitches $\theta_y(t_i)$ by a Gauss-Markov process, such that
\begin{equation}
    \begin{bmatrix} \theta_x(t_i) \\ \theta_y(t_i) \\ \theta_x(t_{i+1}) \\ \theta_y(t_{i+1})\end{bmatrix}\sim \operatorname{Normal}\left( \begin{bmatrix} 0 \\ 0 \\ 0 \\ 0 \end{bmatrix}, \varepsilon^2\begin{bmatrix} 
    1 & 0 & k_{i}^{i+1} & 0 \\ 
    0 & 1 & 0 & k_{i}^{i+1} \\ 
    k_{i}^{i+1} & 0 & 1 & 0 \\ 
    0 & k_{i}^{i+1} & 0 & 1 \end{bmatrix}  \right)
\end{equation}
where $\varepsilon^2$ is the variance of the process, $k_{i}^{i+1} = \exp{\left(-|t_{i+1}-t_i|/l\right)}$ is the correlation of the process between the timesteps $t_i$ and $t_{i+1}$, and $l$ is the lengthscale of the process. The implication of this formulation is that the state of the process at any time $t_i$ is conditionally independent of the state at all previous times apart from the state at the immediately preceding time $t_{i-1}$. We note that this formulation is exactly equivalent to a two-dimensional Brownian motion random walk.

To render the problem more tractable, we approximate our uniform likelihood by a normal likelihood with mean $B(\theta_x,\theta_y)$ and variance $(\Delta_p+\delta_p)^2/12$, where these are simply the mean and variance of the uniform distribution. We assume that there may be some additional excess noise $\sigma^2$ such that the final likelihood is
\begin{align}
    \tilde{B}_i \sim \operatorname{Normal}\bigg(\hat{B}_i + \theta_y(t_i)\cos{\gamma_c} +&s_i\theta_x(t_i)\sin{\gamma_c}, \nonumber \\
    &\frac{1}{12}(\Delta_p+\delta_p)^2+\sigma^2\bigg).
\end{align}
The physical interpretation of the excess noise is that there are either small errors in our model or that there are changes in the roll and pitch occurring on shorter timescales than the time between the epoch observations. By approximating the uniform likelihood by a normal distribution we have reduced the problem to a Gaussian linear model. We note that our Gauss-Markov process in the pitch and roll dimensions are mutually independent a priori, but correlations between them are introduced by the one dimensional measurement of the offset at each timestep.  For fixed values of the parameters $(\varepsilon^2,l,\sigma^2,\delta_c,\delta_p)$, the recursive forward and backward Kalman filter (see Ch. 4 of \citealp{Fraser2008} for an overview) gives analytic bivariate posteriors for the states $\theta_x(t_i)$ and $\theta_y(t_i)$ at each point in time conditionally dependent on the observations at all times. As a by-product, this algorithm also gives an analytic total likelihood $\operatorname{P}(\tilde{B}_{i=1,\dots,N})$ for the model, which we maximised using the Nelder-Mead algorithm \citep{Gao2012} to find the optimal values of the parameters $(\varepsilon^2,l,\sigma^2,\delta_c,\delta_p)$. We fixed these parameters at their maximum likelihood values $\varepsilon = 367.7\;\mathrm{arcsec}$, $l = 0.0061\;\mathrm{days}$, $\sigma = 0.066\;\mathrm{mas}$, $\delta_c = 34.0\;\mathrm{mas}$ and $\delta_p = -0.044\;\mu\mathrm{as}$.

In Fig. \ref{fig:attitude} we illustrate our predictions for the across-scan offsets $\tilde{B}-\hat{B}$ for the two FoVs as a function of time. After exiting the initial Ecliptic Pole Scanning Law, the pattern of across-scan offsets is dominated by periodic oscillations with the same period of $63\;\mathrm{days}$ as the precession period of \gaia. As shown in the bottom middle panel, these are complemented by smaller-scale oscillations with the same period of $6\;\mathrm{hours}$ as the rotation period of \gaia. These two periods describe the nominal rotation of \gaia and thus it makes sense that deviations from the nominal scanning law appear as gyrations at those periods. The bottom left panel illustrates the thinning and widening of the confidence intervals in regions with fewer and more datapoints. We note that out model has very little predictive power in regions more than $l$ away from a datapoint, as clearly shown in the bottom right panel where our model simply linearly interpolates over a gap in the epoch photometry. A substanital improvement on our method would be to model the rotational velocities or accelerations with a Gauss-Markov process, because that would allow the model to meaningfully predict the rotation in these gaps. However, we believe that developing such a model falls outside the scope of this work and that our model is sufficiently accurate for our purposes. We will be accounting for the gaps whenever we predict observations, and so we will never be in the regime of predicting observations without having a nearby epoch observation to anchor the attitude of the scanning law.

In the bottom panel of Fig. \ref{fig:deltahists} we show the across-scan offsets after accounting for our inferred roll and pitch. We now resolve the sharp edges of the gaps between CCDs and can see pixel-to-pixel level variations in the number of associated epoch observations. None of the epoch observations now fall outside of the $0.69\;\mathrm{deg}$ focal plane. We note that differences in the number of observations by the same pixel between the two FoVs will be due to differences in efficiencies between those pixels in the SkyMapper CCDs, which are the only pixels not in common between the two FoVs. We refer the interested reader to \citet{Crowley2016} for an in-depth analysis of the pixel-to-pixel and CCD-to-CCD variations in \gaia's camera.

We identified the posterior of our model for the roll and pitch of \gaia at the timepoints of the DPAC-published nominal scanning law by inserting those times into the Kalman filter calculation as times without associated observations. The rotation described by the matrix in Eq. \ref{eq:rotmatrix} implies a change in the location of the preceeding and following FoVs on the sky, and thus we are able to derive a much more accurate scanning law for the period covered by \gaia DR2, reducing the across-scan motion error from $30\;\mathrm{arcsec}$ down to $0.066\;\mathrm{arcsec}$. We stress that our model has only constrained the scanning law in the across-scan direction and that it is still possible for the fields of view to be wrong by the quoted $30\;\mathrm{arcsec}$ in the along-scan direction, which could lead to predicted times of observations being wrong by as much as $500\;\mathrm{ms}$.

In this Appendix we have used the epoch photometry of the 550,737 variable stars published in \gaia DR2 to derive a more accurate scanning law than the nominal scanning law published by DPAC. We note that the \gaia DPAC do have a much more accurate scanning law which was derived as a by-product of the astrometric pipeline, however they have not at this point published it. Therefore, our determination is the most accurate determination of the \gaia scanning law in the public domain. We have published our scanning law as an electronic file on the Harvard Dataverse (\url{https://doi.org/10.7910/DVN/OFRA78}). An example of the file format is given in Tab. \ref{tab:scanning law}, which we note is the exact same format used by the \gaia DPAC for their nominal scanning law file.

\section{Dim stars}
\label{sec:dimstars}

\begin{figure}
	\centering
	\includegraphics[width=1.\linewidth,trim=0 0 0 0, clip]{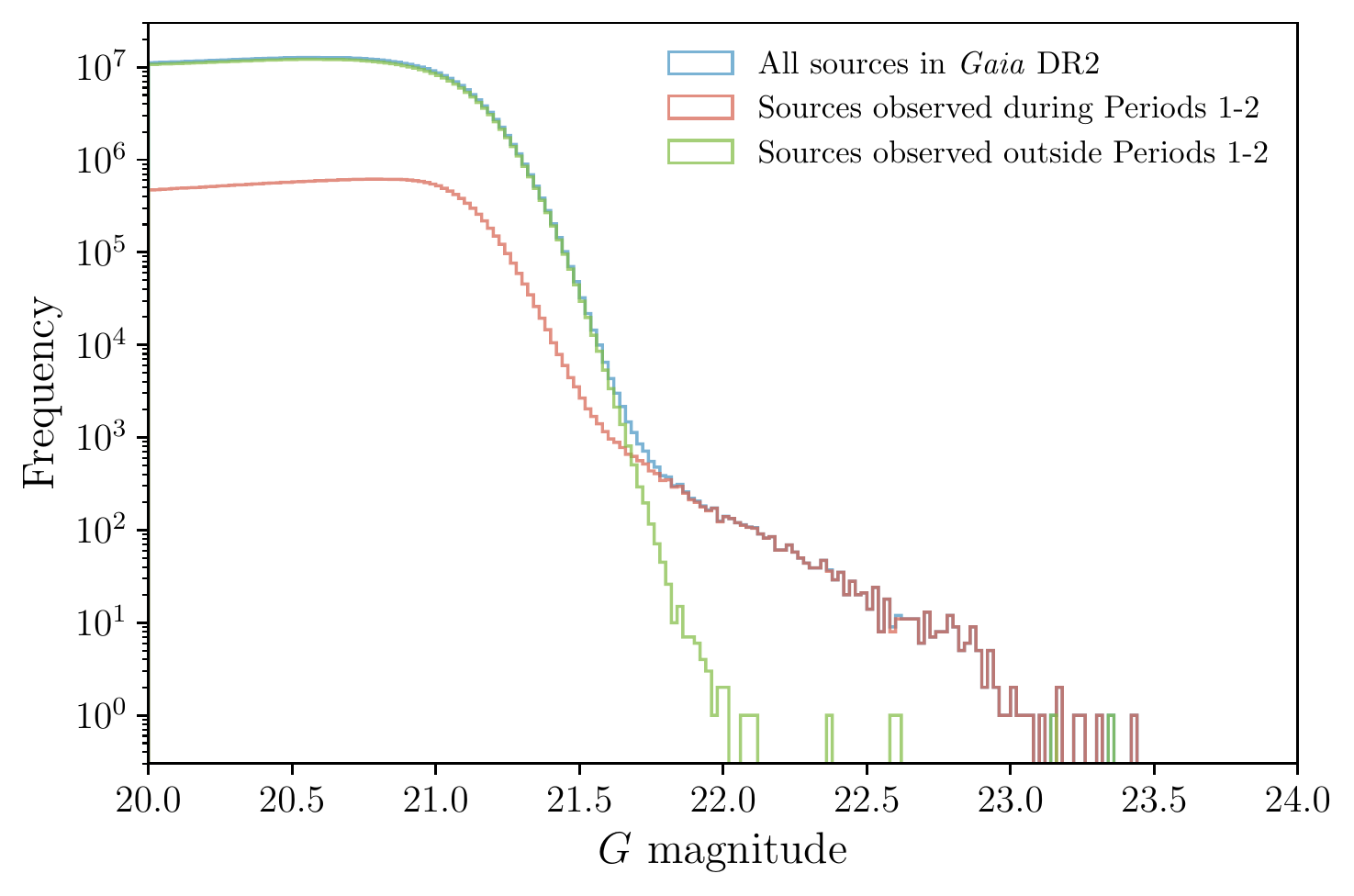}
	a) Faint magnitude distribution of stars in \gaia DR2.\vspace{0.2cm}
	
	\includegraphics[width=1.\linewidth,trim=10 10 10 10, clip]{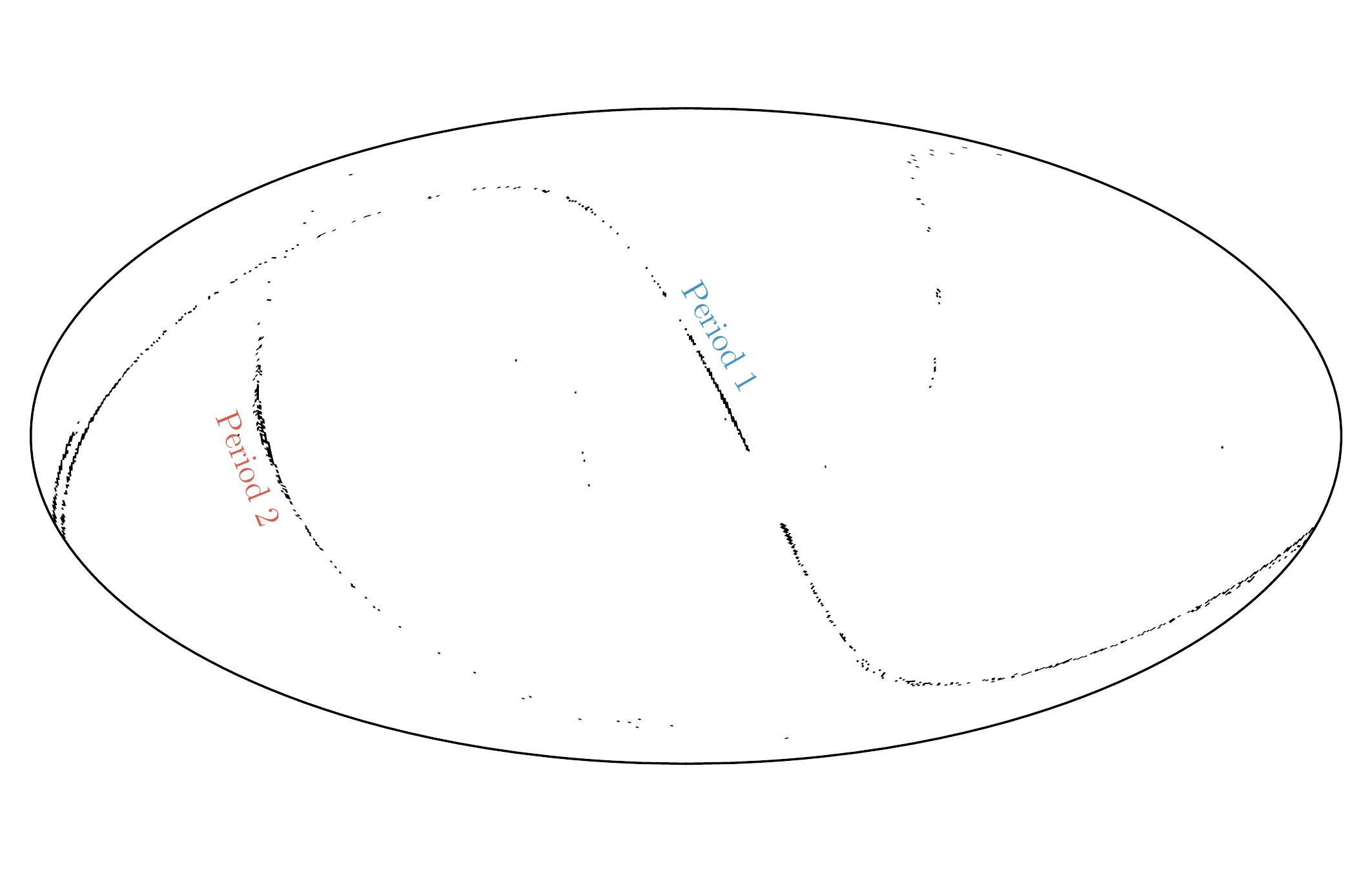}
	b) Galactic distribution of stars with $G>21.7$.\vspace{0.2cm}
	
	\includegraphics[width=1.\linewidth,trim=10 10 10 10, clip]{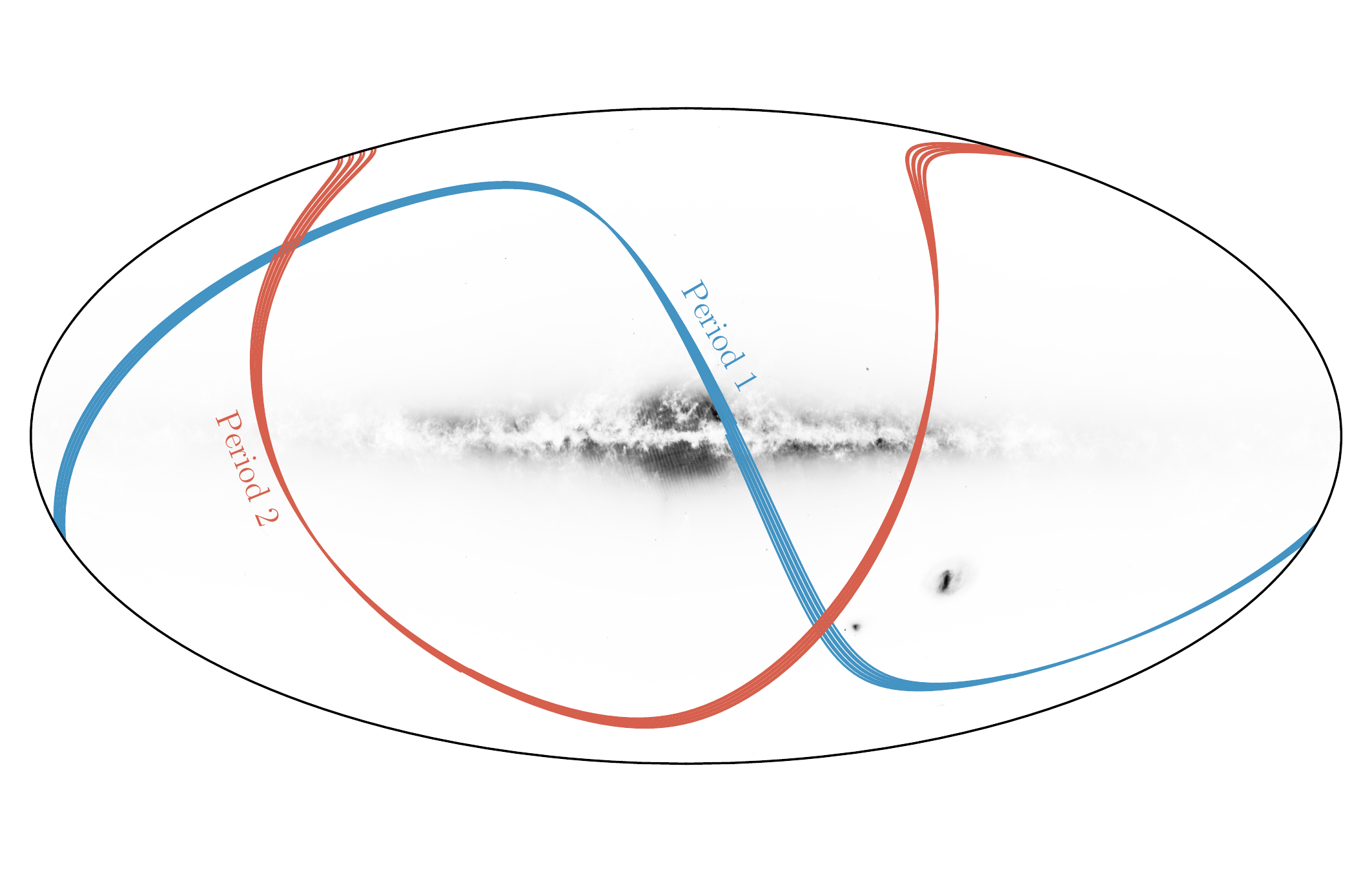}
	c) \gaia scans responsible for the faintest sources.
	
	\caption{There are sources in \gaia with reported mean $G$-band magnitudes as faint as $G=23.5$. We identified that an overwhelming majority of these sources were observed during two narrow time windows and so are likely due to missing calibration data packets preventing an accurate magnitude determination. \textbf{Top:} Faint magnitude distribution of all sources in \gaia DR2 (blue), of sources predicted to have been observed at least once during the two time periods identified in the main text (red), and of those sources with no predicted observation during those periods. Almost all of the extremely faint sources can be traced to those time periods. \textbf{Middle:} Galactic coordinate distribution of stars fainter than $G>22$. The dynamic range of this plot has been limited to emphasise the low number features. Almost all of the stars lie along two narrow tracks which correspond to the strips of sky observed by \gaia during small time windows. \textbf{Bottom:} We identified the time windows corresponding to each of the two strips and show where \gaia was looking during these windows. A HEALPix map of all \gaia DR2 sources is shown in the background to contextualise the scans relative to the Galaxy.}
	\label{fig:dimstars}
\end{figure}

99.865\% of the sources in \gaia DR2 are brighter than $G=21.3$ \citep{Gaia2018}, but there is a tail of fainter sources out to $G=23.5$ which we show in the top panel of Fig. \ref{fig:dimstars}. These sources are likely to be spurious because \gaia is not sufficiently sensitive to detect sources this faint in the short time in which sources transit the focal plane. The drop-off in the number of sources at each magnitude shows a change in behaviour at around $G=21.7$ and we conjecture that the magnitudes of most of the sources fainter than $G=22$ are likely to be spurious. We show the on-sky distribution of the 1,869 stars fainter than $G=22$ in Galactic coordinates in the middle panel of Fig. \ref{fig:dimstars}. Almost all of these sources lie along two narrow strips and thus can be attributed to specific periods of the \gaia scanning law. Using a new tool developed in Holl et al. (in prep.) for this use-case, we identified two rough time ranges which we label Period 1 ($\mathrm{OBMT}=1388{-}1392\;\mathrm{rev}$) and Period 2 ($\mathrm{OBMT}=2211{-}2215\;\mathrm{rev}$), and we illustrate where \gaia was scanning during these periods in the bottom panel of Fig \ref{fig:dimstars}. Curiously, Period 1 aligns with the drop in the colour photometry efficiency mentioned in Sec. \ref{sec:comparison}.

The cause of the miscalibration in these periods was a break of communication with the spacecraft which resulted in the loss of some scientific data and the delay in transmission of critical auxiliary science data (ASD) which was then not used in \gaia DR2 processing (\gaia Helpdesk, private communication). There were several stray light peaks within the down-time which were therefore missed by the calibration and instead interpolated over. This directly affected measured fluxes of many of the sources observed in the range OBMT=1389.2-1391.7, which resulted in $G_\mathrm{BP}$ and $G_\mathrm{RP}$ observations not being included in the epoch photometry as seen in Fig.~\ref{fig:slowdown}.
\gaia's photometric calibration occurs in one day time intervals which in this case corresponds to the interval $\mathrm{OBMT}=1388.0{-}1392.0\;\mathrm{rev}$. Those observations within the range $\mathrm{OBMT}=1389.2{-}1391.7\;\mathrm{rev}$ which were directly affected received overestimated fluxes due to the unobserved stray-light peaks. Observations within the calibration interval but not directly affected by the down-time received underestimated fluxes due to the calibration process effectively averaging out the observed error. It is these stars which appear as extremely dim sources in Period 1.
Period 2 is also likely a result of background interpolation issues however we do not have an explanation of the exact cause of this particular event.


\bsp	
\label{lastpage}
\end{document}